\documentclass[journal]{IEEEtran}
\usepackage{latexsym,amssymb,amsmath,graphicx,epsfig,bbm,float,epsf,graphics}
\usepackage{algorithm}
\usepackage{algpseudocode}
\usepackage{tabularx}
\usepackage{url}
\usepackage{multicol}
\usepackage{amsmath}
\usepackage{subcaption}
\usepackage{relsize}
\usepackage{tcolorbox}
\usepackage{mathtools}
\usepackage{enumitem}
\usepackage{multirow}
\usepackage{booktabs}
\usepackage[font=small,labelfont=bf]{caption}
\usepackage{comment}
\usepackage{cite}

\newcolumntype{C}[1]{>{\centering\let\newline\\\arraybackslash\hspace{0pt}}m{#1}}
\renewcommand{\vec}[1]{\mbox{\boldmath${#1}$}}
\DeclareMathOperator*{\argmin}{arg\,min}

\begin{document}
%
\title{Bayesian Sparse Blind Deconvolution Using MCMC Methods Based on Normal-Inverse-Gamma Prior}

\author{Burak C. Civek and
        Emre Ertin
\thanks{B. C. Civek and E. Ertin are with the Department
of Electrical and Computer Engineering, The Ohio State University, Columbus, OH, 43210, USA. Contact e-mail: civek.1@osu.edu}}

\maketitle

\begin{abstract}
Bayesian estimation methods for sparse blind deconvolution problems conventionally employ Bernoulli-Gaussian (BG) prior for modeling sparse sequences and utilize Markov Chain Monte Carlo (MCMC) methods for the estimation of unknowns. However, the discrete nature of the BG model creates computational bottlenecks, preventing efficient exploration of the probability space even with the recently proposed enhanced sampler schemes. To address this issue, we propose an alternative MCMC method by modeling the sparse sequences using the Normal-Inverse-Gamma (NIG) prior. We derive effective Gibbs samplers for this prior and illustrate that the computational burden associated with the BG model can be eliminated by transferring the problem into a completely continuous-valued framework. In addition to sparsity, we also incorporate time and frequency domain constraints on the convolving sequences. We demonstrate the effectiveness of the proposed methods via extensive simulations and characterize computational gains relative to the existing methods that utilize BG modeling.
\end{abstract}


\section{Introduction}
\IEEEPARstart{T}{he} problem of reconstructing an unknown signal from an observation, where the observation is modeled as the filtered version of the unknown signal with a given distortion filter, is known as the deconvolution problem. When the impulse response of the distortion filter is not known and to be estimated along with the unknown signal, it evolves into the Blind Deconvolution (BD) problem. This problem setting emerges in a variety of real-life applications including, but not limited to, image deblurring \cite{LEdmund}, seismic exploration \cite{QCheng,JMendel}, digital communication \cite{GXu,EMoulines}, and biomedical signal reconstruction \cite{JGao,BCivek}. The class of BD problems is inherently ill-posed and suffers from the identifiability issues without the prior knowledge about the convolving signals \cite{SChoudhary}. This is due to the fact that many distinct signal pairs can yield the same observation. To overcome this issue, several different assumptions are made on the signal pairs in an attempt to constrain the solution space. Typical examples of these include restricted support in frequency or time domain \cite{JGao,BCivek}, availability of a sparse representation \cite{CBilen,YLi,LWang}, or existence of a generative subspace \cite{EMoulines,AAhmed}.\par

Solution of BD problems has been extensively studied under both deterministic and probabilistic frameworks. Deterministic approaches typically construct a constrained optimization problem, where the goal is to minimize an objective function corresponding to a likelihood term that measures the quality of fit. The constraints on the variables are imposed either explicitly or by augmenting the objective function. Because a closed form solution is usually not available, gradient based search algorithms are employed to converge to a local minimum that minimizes the objective function. However, due to the structure of the constraints, the constructed optimization problem usually has a non-convex formulation, and therefore, the successful recovery relies on finding a good initial estimate \cite{YLi}. Even though there exist convex formulations, which eliminate the effect of initialization, their ability to impose a variety of constraints simultaneously, e.g., frequency and time domain constraints at the same time, is limited \cite{BCivek,CBilen,LWang}. Notably, the lifting approach proposed in \cite{AAhmed}, and its extension to sparse sequences \cite{SLing}, enable imposing time/frequency constraints jointly by assuming the existence of a generative subspace. However, the number of available measurements is limited when the impulse response of the distortion filter has a bandlimited structure, which might prevent successful recovery with lifting approaches. \par 

Probabilistic approaches, on the other hand, provide effective alternatives by modeling the unknown quantities as random variables and producing estimates based on the posterior distribution. Appropriate prior distributions are assigned to the unknown variables in order to increase the likelihood of points in the probability space that satisfy the constraints. Similar to deterministic models, analytic solution is not tractable in many cases, where the solution is obtained through numerical iterative methods such as Markov Chain Monte Carlo (MCMC) \cite{CRobert}. Different from gradient-based searches, MCMC methods do not rely on gradient information, which reduces the effect of initialization on the convergence behavior \cite{gilks1995markov}. This enables probabilistic schemes to incorporate more complex constraints simultaneously without suffering from the highly non-convex structure of the posteriors shaped by these constraints.\par

In this paper, we adopt the probabilistic framework and present computationally efficient Bayesian methods for the solution of the regularized BD problem, in which the desired signal is a sparse sequence and the unknown impulse response of the distortion filter is time and/or bandlimited. This problem setting is of great importance due to its applicability to a wide range of scenarios. In many cases, even if the desired signal itself is not sparse, it is possible to find a sparse representation in a suitable transform domain. Moreover, physical realizations of real-world systems are well modeled by exactly or approximately bandlimited system responses with finite impulse responses.\par

Bayesian BD methods conventionally employ a Bernoulli-Gaussian (BG) prior for modeling the sparse sequences \cite{MLavielle,CSoussen}. According to the BG model, the sparse sequence is represented by a Bernoulli distributed binary latent sequence, indicating the nonzero positions, and an amplitude sequence, representing the amplitudes corresponding to those positions under the Gaussian law. The posterior distribution constructed in this manner has a discrete structure, prohibiting efficient analytical solutions and leading to iterative approaches for making inference on the posterior. Bayesian methods typically use the MCMC simulations, whose potential in the sparse BD literature was presented by the pioneering work of Cheng et al. \cite{QCheng}. The MCMC methods provide powerful iterative tools, such as Gibbs sampler, for Bayesian inference problems, even when the number of unknowns is quite high \cite{SGeman}. The Gibbs sampler performs quite well with a significantly fast mixing rate, especially if the sampled variables are independent \cite{SGeman}. However, the mixing performance degrades considerably when there are strong dependencies between variables \cite{SBourguignon,DGe}. \par 

It has been shown that the original Gibbs sampler constructed based on BG prior for sparsity causes implicit dependencies between the consecutive sampling steps of the latent indicator sequence, which prohibits the sampler to explore the probability space efficiently, causing it to get stuck on a local optimum for a long time \cite{SBourguignon}. In order to eliminate this deficiency, more efficient sampling methods were proposed in \cite{SBourguignon} and \cite{CChi}, accounting for the statistical dependence between the neighboring  variables in the indicator sequence. An alternative idea, presented in \cite{DGe}, is to use the blocked Gibbs sampler scheme, which was first introduced in \cite{JLiu}, that enables sampling adjacent variables jointly. Despite the considerable improvement in the convergence rate, their application is usually limited to very short blocks due to exponentially increasing computational complexity with the block length. Another attempt, also presented in \cite{DGe}, to improve the mixing rate was to integrate the partially collapsed Gibbs (PCG) sampler \cite{DDyk}, a generalization of the block Gibbs samplers, into the BG model. PCG samplers explore the probability space more effectively by means of marginalization and trimming operations \cite{DDyk}. However, although the PCG sampling scheme outperforms the blocked sampler schemes in terms of the number of iterations needed for convergence, iterations are considerably more costly \cite{DGe}.\par

The problems stated above are mainly due to the discrete nature of the BG model, which creates a computational bottleneck even for the highly efficient sampling schemes. In order to address this deficiency, we present alternative MCMC methods which utilize the Normal-Inverse-Gamma (NIG) prior for modeling sparsity. Unlike the BG model, the sparse sequence is modeled by a zero-mean multivariate Gaussian distribution with an unknown diagonal covariance matrix, where the individual variances marginally follow an Inverse-Gamma (IG) distribution. Therefore, the marginal distribution for the sparse sequence corresponds to a multivariate $t$-distribution, whose heavy tails encourage sparsity \cite{YildirimS,MohammadA,YangL}. With this setting, the problem is transformed into a completely continuous valued domain, i.e., all unknowns including the latent variables are continuous valued. We first present a classical Gibbs sampling approach that exploits the continuous valued structure of the problem provided by the NIG prior. We then propose a PCG based sampling scheme to further enhance the convergence behavior by accounting for statistical dependency. The performance gains achieved by the proposed methods are illustrated through a variety of simulations. To the best of our knowledge, this paper presents the first application of the MCMC methods based on the NIG prior to the Bayesian sparse BD problem with time/frequency domain constraints. The proposed methods achieve state-of-the-art performance of the previously proposed blocked \cite{DGe} Gibbs samplers with BG prior at a fraction of their computational cost.\par

The paper is organized as follows. We first introduce the problem setting and define the prior distributions associated with each variable in Section \ref{problem_setting}. Then, in Section \ref{proposed_samplers}, we present the proposed MCMC methods for the solution of sparse BD problems, followed by their validation through numerical experiments in Section \ref{simulations}. We finalize the discussion with the concluding remarks in Section \ref{conclusion}.

\section{Problem Statement}\label{problem_setting}
We consider a real-valued time-domain observation sequence $y_n$ of length $N$, which is modeled by linear convolution of two finite sequences, a relatively shorter pulse sequence $h_n$ of length $T \ll N$, modeling the impulse response of the distortion filter, and a sparse sequence $x_n$ of length $K < N$. The output of the convolution is then corrupted by additive noise $v_n$, yielding the following measurement model
\begin{equation}\label{conv_model}
    y_n = \sum_{k = 0}^{T-1}h_k x_{n-k} + v_n\; \text{ for } n = 0,...,N-1,
\end{equation}
where $x_n = 0$ for any $n \notin \{0,\hdots,K-1\}$. Here, we consider the scenario where both the pulse shape $h_n$ and the sparse sequence $x_n$ are unknown and to be estimated from the observation $y_n$. This problem is well-known as the blind deconvolution problem and usually ill-posed if there is no prior information about the pair of convolving sequences. Here, we focus on the setting where the problem is regularized by sparsity and time-frequency domain constraints, which enables the successful recovery of both sequences. \par 

For notational convenience, we define the vector valued variables $\vec{y},\vec{v} \in \mathbbm{R}^N$, $\vec{x} \in \mathbbm{R}^K$, and $\vec{h} \in \mathbbm{R}^T$, and rewrite (\ref{conv_model}) as
\begin{equation}\label{conv_model2}
\vec{y} = \vec{H}\vec{x}+\vec{v},
\end{equation}
where $\vec{H} \in \mathbb{R}^{N \times K}$ is the Toeplitz matrix of $\vec{h}$. Here, the first column and row of $\vec{H}$ are given as $[h_0,h_1,...,h_{T-1},0,...,0]^T$ and $[h_0,0,...,0]$ respectively. Defining $\vec{X}\in \mathbb{R}^{N \times T}$ as the Toeplitz matrix having $[x_0,x_1,...,x_{K-1},0,\hdots,0]^T$ as its first column and $[x_0,0,...,0]$ as its first row, we can alternatively rewrite (\ref{conv_model2}) as $\vec{y} = \vec{X}\vec{h}+\vec{v}$. We will make use of both of these formulations
in the rest of the paper.

\subsection{Prior Distributions}\label{section_prior_dist}
Following the Bayesian framework, the unknowns are modeled as random variables with specific prior distributions. This section provides a complete description of the prior distributions assigned to each variable. We begin with review of the BG distribution, which is the conventional prior used in Bayesian settings for sparse sequences. We then introduce the NIG model, which is a \textit{soft} alternative to the BG model promoting sparsity of $\vec{x}$ and will constitute the basis of our proposed estimator in the next section. \par 

\subsubsection{Bernoulli-Gaussian Prior for Sparsity}
The BG model introduces a latent binary sequence $\vec{s} = [s_0,s_1,\hdots,s_{K-1}]^T$ with $s_n \in \{0,1\}$ and defines the conditional distribution of $x_n$ given $s_n$ as
\begin{equation}\label{conditional_prior_x_given_s}
    p(x_n|s_n) =
    \begin{cases}
    \delta(x_n) &\text{if} \;\; s_n = 0\\
    \mathcal{N}(x_n;0,\sigma_{x}^2) &\text{if} \;\; s_n = 1
    \end{cases},
\end{equation}
where $\delta(\cdot)$ is the Dirac delta function and $\mathcal{N}(\cdot;\mu,\sigma^2)$ denotes the Gaussian distribution with mean $\mu$ and variance $\sigma^2$. Assuming $s_n$ are independent and identically distributed (i.i.d.) according to a Bernoulli distribution with parameter $\pi_0 = P(s_n = 0)$, the prior for $\vec{s}$ takes the form
\begin{equation}\label{prior_s}
    p(\vec{s}) = \prod_{n=0}^{K-1}p(s_n) = \binom{K}{|\mathcal{K}_1|}\pi_0^{|\mathcal{K}_1|}(1-\pi_0)^{K-|\mathcal{K}_1|},
\end{equation}
where $\mathcal{K}_1$ denotes the set of indices giving the locations of 1's and $|\mathcal{K}_1|$ represents the cardinality of $\mathcal{K}_1$, i.e., $|\mathcal{K}_1| = \sum s_n$. Assuming different pairs of $(x_n,s_n)$ are statistically independent, the joint distribution of $\vec{x}$ and $\vec{s}$ becomes
\begin{equation}\label{joint_prior_x_and_s}
    p(\vec{x},\vec{s}) = p(\vec{s})\prod_{n \in \mathcal{K}_1}\mathcal{N}(x_n;0,\sigma_{x}^2)\prod_{n\in\mathcal{K}_0}\delta(x_n),
\end{equation}
where $\mathcal{K}_0$ is the complement of $\mathcal{K}_1$ in $\{0,1,...,K-1\}$. The parameter $\pi_0$ reflects our \textit{a priori} knowledge about the expected rate of 1's in $\vec{s}$, limiting the total number of nonzero entries in $\vec{x}$, and hence, leading to a sparse sequence.
\subsubsection{The Normal-Inverse-Gamma Prior for Sparsity}
Instead of introducing a latent binary sequence, which forms a discrete probability space, the sparsity can also be imposed by a diagonal covariance matrix $\vec{\Sigma}_x = \text{diag}(\vec{\sigma}_x^2)$, where $\vec{\sigma}_x^2 = [\sigma_{x_0}^2,\sigma_{x_1}^2,\hdots,\sigma_{x_{K-1}}^2]^T$. Here, $\vec{\sigma}_x^2$ consists of continuous valued unknown variances of each element in $\vec{x}$. Under a zero-mean multivariate Gaussian law with the covariance matrix $\vec{\Sigma}_x$, the conditional prior distribution of $\vec{x}$ given $\vec{\sigma}_x^2$ becomes 
\begin{equation}\label{conditional_prior_x_given_sigma_x}
    p(\vec{x}|\vec{\sigma}_x^2) = \mathcal{N}(\vec{x};\vec{0},\vec{\Sigma}_x),
\end{equation}
where $\mathcal{N}(\cdot;\vec{\mu},\vec{\Sigma})$ represents the multivariate Gaussian distribution with mean $\vec{\mu}$ and covariance $\vec{\Sigma}$. The idea is that an individual element $x_n$ can be made arbitrarily small by setting a sufficiently low variance $\sigma_{x_n}^2$. Therefore, the unknown variance vector $\vec{\sigma}_x^2$ is also assumed to be a random sequence and to be estimated along with $\vec{h}$ and $\vec{x}$. \par

\begin{figure}[t!]
    \centering
    \includegraphics[width=1\linewidth]{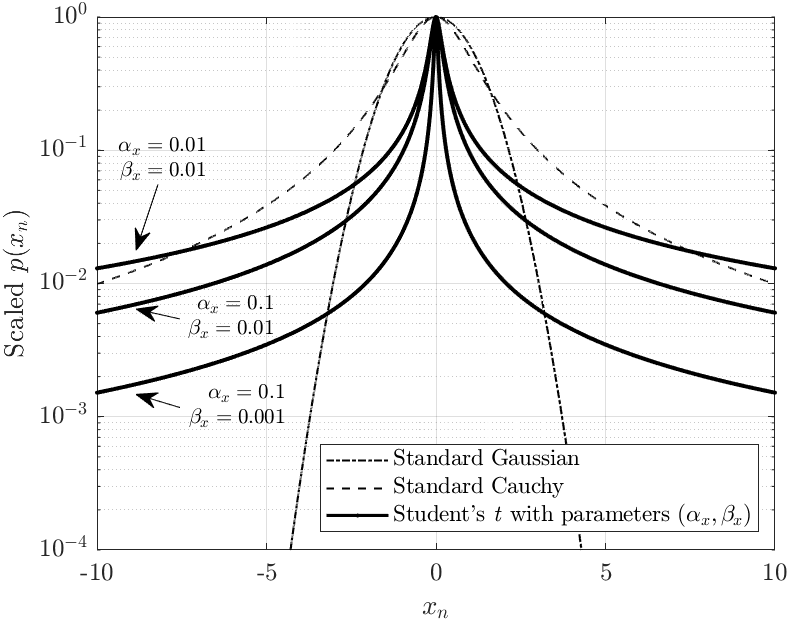}
    \caption{Marginal prior distribution of $x_n$ in logarithmic scale for different parameter values $(\alpha_x,\beta_x)$. Standard Gaussian and Cauchy distributions are included for comparison. All densities are scaled such that the maximum is 1. \label{marginal_prior_of_x}}
    \vspace{-5mm}
\end{figure}

We assign i.i.d. IG\footnote{Please see the supplement for the explicit definition of IG distribution.} prior on $\vec{\sigma}_x^2$ with shape and scale parameters $\alpha_x$ and $\beta_x$
\begin{equation}\label{prior_sigma_x}
    p(\vec{\sigma}_x^2|\alpha_x,\beta_x) = \prod_{n = 0}^{K-1}\mathcal{IG}(\sigma_{x_n}^2;\alpha_x,\beta_x).
\end{equation}
The IG distribution is conjugate prior for the unknown variance of the Gaussian distribution, which enables analytical calculation of the posterior. Moreover, note that the marginal distribution of any element $x_n$ becomes
\begin{equation}\label{marginal_prior_x}
\begin{split}
    p(x_n) &= \int p(x_n|\sigma_{x_n}^2)p(\sigma_{x_n}^2)d\sigma_{x_n}^2 \\
    &= \dfrac{\beta_x^{\alpha_x}}{\sqrt{2\pi}\Gamma(\alpha_x)}\dfrac{\Gamma(\alpha_x + 0.5)}{(0.5x_n^2 + \beta_x)^{\alpha_x + 0.5}},
\end{split}
\end{equation}
which corresponds to a generalized Student's \textit{t}-distribution with degree of freedom $2\alpha_x$ and scale $\beta_x/\alpha_x$. As shown in Fig. \ref{marginal_prior_of_x}, appropriate selection of parameters $(\alpha_x,\beta_x)$ leads to a family of distributions that are highly concentrated around zero with significantly heavier tails compared to standard Gaussian distribution, justifying the use of NIG model for sparse sequences as an alternative to BG model. However, as opposed to the simple interpretation of $\pi_0$ within the BG model, the effect of parameters $\alpha_x$ and $\beta_x$ is complicated, preventing the availability of prior estimates beforehand. Therefore, we construct a hierarchical Bayesian model and learn the prior distribution parameters $\alpha_x$ and $\beta_x$ from the measurement as well. Assuming no prior information on $\alpha_x$ and $\beta_x$, we assign Jeffreys prior, given by $p(\alpha_x) \propto 1/\alpha_x$ and $p(\beta_x) \propto 1/\beta_x$, which forms an improper prior exhibiting non-informative structure. Here, $\propto$ denotes the proportionality. We should note that the effect of prior distributions $p(\alpha_x)$ and $p(\beta_x)$ on the posterior will indeed be dominated by $p(\vec{x}|\vec{\sigma}_x^2)$ and $p(\vec{\sigma}_x^2|\alpha_x,\beta_x)$ as the length of the sequence $K$ increases. Hence, the constructed hierarchical model is not sensitive to the selection of prior distributions for the variables $\alpha_x$ and $\beta_x$.

\subsubsection{Prior for Short Pulse Sequence}
Time and frequency domain constraints for pulse sequences are widely used in blind deconvolution framework to regularize the problem \cite{JGao,BCivek,GKail}. One common practice is to assume that the pulse sequence belongs to a known subspace, i.e., $\vec{h} = \vec{A}\vec{\gamma}$, where $\vec{A} \in \mathbbm{R}^{T \times L}$ represents a lower dimensional subspace with $L \leq T$ and $\vec{\gamma} \in \mathbbm{R}^L$ represents the unknown orientation of $\vec{h}$ in the subspace, which is to be estimated. The duration of $\vec{h}$ in time domain is explicitly enforced by the dimension of $\vec{A}$, and the frequency domain restrictions can be applied by constructing $\vec{A}$ using the first $L$ sequence of either Discrete Prolate Spheroidal (DPS) Sequences or Hermite Functions \cite{FHlawatsch}. Whenever there is no specific frequency domain restriction, $\vec{A}$ can be set as identity, i.e., $\vec{A} = \vec{I}$. We should note that due to the scaling ambiguity inherent in BD problems, the scale of the pulse sequence $\vec{h}$ must be restricted. This can be achieved by assigning an appropriate prior on $\vec{\gamma}$. Therefore, we assign a zero-mean i.i.d. Gaussian distribution with variance $\sigma_{\gamma}^2$, i.e.,
\begin{equation}\label{prior_gamma}
    p(\vec{\gamma}) = \mathcal{N}(\vec{\gamma};\vec{0},\sigma_{\gamma}^2\vec{I}).
\end{equation}
Here, the variance $\sigma_{\gamma}^2$ is a fixed hyperparameter to avoid scaling ambiguity.
\subsubsection{Prior for Noise Variance}
We assume that measurements are corrupted by additive white Gaussian noise with unknown variance $\sigma_v^2$. Assuming no prior information, similar to the other parameters, we assign Jeffreys prior, i.e., $p(\sigma_v^2) \propto 1/\sigma_v^2$.

\subsection{Estimation Problem}
The estimation problem consists of estimating the actual variables of interest $\vec{x}$ and $\vec{\gamma}$, along with the noise variance $\sigma_v^2$, the latent variables $\vec{\sigma}_x^2$, and the corresponding prior distribution parameters $\alpha_x$, $\beta_x$ from the given measurement $\vec{y}$. Given the prior distributions for all variables, the posterior distribution follows
\begin{equation}\label{posterior}
    p(\vec{\theta}|\vec{y}) \propto p(\vec{y}|\vec{\theta})p(\vec{\theta}),
\end{equation}
where we set $\vec{\theta} = [\vec{x},\vec{\sigma}_x^2,\vec{\gamma},\sigma_v^2,\alpha_x,\beta_x]$ for more compact notation. Assuming all variables are statistically independent, the prior distribution $p(\vec{\theta})$ is given by
\begin{equation}\label{prior}
    p(\vec{\theta}) = p(\vec{x}|\vec{\sigma}_x^2)p(\vec{\sigma}_x^2|\alpha_x,\beta_x)p(\alpha_x)p(\beta_x)p(\vec{\gamma})p(\sigma_v^2),
\end{equation}
where the expressions for the right hand side are given in Section \ref{section_prior_dist}. Assumed Gaussian noise model yields the following likelihood term
\begin{equation}\label{likelihood}
    p(\vec{y}|\vec{\theta}) \propto \bigg(\dfrac{1}{\sigma_v^2}\bigg)^{N/2}\exp\bigg(-\dfrac{\|\vec{y} - \vec{X}\vec{A}\vec{\gamma}\|^2}{2\sigma_v^2}\bigg),
\end{equation}
where $\|\cdot\|$ denotes $\ell_2$ norm of a vector. Note that the likelihood term is only a function of $\vec{x}$, $\vec{\gamma}$ and $\sigma_v^2$.\par 
For estimation of the variables, we consider the minimum mean-square-error (MMSE) estimator, i.e.,
\begin{equation}\label{mmse_estimator}
    \vec{\theta}^{*} = E[\vec{\theta}|\vec{y}],
\end{equation}
which is equivalent to the expected value of the posterior distribution given in (\ref{posterior}). However, explicit calculation of (\ref{mmse_estimator}) is not possible since it requires analytically intractable integrations. This leads us to an approximate solution, which can be obtained using MCMC simulations. MCMC methods are widely used in Bayesian inference problems, including blind deconvolution literature, when the posterior distribution is too complicated to obtain exact analytical solutions \cite{CRobert,gilks1995markov}. The first step of MCMC methods consists of generating a set of, say, $J$ random samples, denoted by $\{\vec{\theta}^{(i)}\}_{i=1}^{J}$, with $\vec{\theta}^{(i)} = [\vec{x}^{(i)},\vec{\sigma}_x^{2(i)},\vec{\gamma}^{(i)},\sigma_v^{2(i)},\alpha_x^{(i)},\beta_x^{(i)}]$ being the $i^{th}$ sample, from the posterior distribution using an appropriate sampler. Once the sampling process is completed, the actual solution of (\ref{mmse_estimator}) is approximated by the sample mean, i.e., 
\begin{equation}\label{MCMC_approximation}
    \vec{\theta}^{*} \simeq \dfrac{1}{J-J^{\prime}}\sum_{i=J^{\prime}+1}^{J}\vec{\theta}^{(i)},
\end{equation}
where the first $J^{\prime}$ samples are discarded as part of the burn-in process. Here, the crucial part is to have an effective sampler, which can converge to the true target distribution quickly. To this end, in the next section, we construct the proposed valid sampling schemes to be used within MCMC simulations.
\section{Proposed Samplers for Sparse Blind Deconvolution}\label{proposed_samplers}
Samplers are essential components of MCMC methods for complex target distributions. They construct Markov Chains whose stationary distributions converge to the target distribution (which corresponds to the posterior distribution (\ref{posterior}) in our case) in the long run \cite{WGilks2}. The effectiveness of the sampler is directly associated with the mixing rate, which represents how fast the stationary distribution is achieved. In this work, we employ both Gibbs and PCG sampling schemes with improved mixing rates.\par

In this section, we first briefly review the idea of classical Gibbs sampling for multivariate distributions, followed by the construction of our proposed classical Gibbs sampler, which utilizes the alternative NIG model for sparsity. Then, we propose a PCG based sampler to improve upon the classical Gibbs sampler to obtain a faster mixing rate with a slightly increased computational complexity. 
\subsection{Gibbs Sampler}
Let $\pi(\vec{\theta})$ denote the target distribution that we want to sample from, with $\vec{\theta}$ being a vector valued random variable of arbitrary length, say $M$, i.e., $\vec{\theta} = [\theta_0,\theta_1,\hdots,\theta_{M-1}]^T$. When direct sampling from $\pi(\vec{\theta})$ is not feasible, the idea of Gibbs sampling suggests that we can sample each of the scalar variables, $\theta_m$, in turn from their conditional distributions with all other variables are fixed at their current values. Hence, in the $i^{th}$ iteration, $\theta_m^{(i)}$ is obtained by sampling from $p(\theta_m^{(i)}|\theta_{m^-}^{(i)},\theta_{m^+}^{(i-1)})$, where $m^-$ and $m^+$ represents the indices $\{1,\hdots,m-1\}$ and $\{m+1,\hdots,M\}$ respectively. One iteration of Gibbs sampling is completed once every single variable is updated. All variables are initialized, usually by sampling from their prior distributions, before the first iteration, which has a strong effect on the first few realizations. In order to eliminate this effect, a burn-in process is incorporated, where the realizations generated until convergence to the target distribution are discarded. \par
It is also useful to note that this process is not restricted to sampling a single scalar variable at a time. An extension of the Gibbs sampler, called blocked Gibbs sampler, allows sampling blocks of variables at one step through their joint distribution conditioned on others. It helps to achieve considerably improved convergence rates compared to sampling a scalar valued variable at a time by reducing the autocorrelation between the successive samples. This is especially useful when applied to variables with strong dependencies.
\begin{table}
\centering
\normalsize
\caption{Proposed Classical Gibbs Sampler}
\vspace{-2mm}
\label{table1}
\renewcommand\arraystretch{1.2}
\begin{tabular}{|m{0.4\textwidth}|}
\hline
Step 1. Sample $\alpha_x$ from $p(\alpha_x|\vec{\sigma}_x^2,\beta_x)$  \\
Step 2. Sample $\beta_x$ from $p(\beta_x|\vec{\sigma}_x^2,\alpha_x)$  \\
Step 3. Sample $\vec{\sigma}_x^2$ from $p(\vec{\sigma}_x^2|\vec{x},\alpha_x,\beta_x)$  \\
Step 4. Sample $\vec{x}$ from $p(\vec{x}|\vec{y},\vec{\sigma}_x^2,\vec{\gamma},\sigma_v^2)$ \\
Step 5. Sample $\vec{\gamma}$ from $p(\vec{\gamma}|\vec{y},\vec{x},\sigma_v^2)$ \\
Step 6. Sample $\sigma_v^2$ from $p(\sigma_v^2|\vec{y},\vec{x},\vec{\gamma})$ \\
\hline
\end{tabular}
\vspace{-5mm}
\end{table}
\subsection{The Proposed Classical Gibbs Sampler}\label{proposed_classical_gibbs_sampler}
We begin with constructing the classical Gibbs sampler based on the selected prior distributions. Then, in the next section, we propose an alternative sampling scheme that introduces additional intermediate steps to enhance the mixing rate. One iteration of our classical Gibbs sampler is described in Table \ref{table1}. In each step, we sample a variable from its full conditional posterior distribution. The reason for dropping some of the variables from the conditions is not because of marginalization, but due to conditional independence. For instance, in Step 1, we have $p(\alpha_x|\vec{y},\vec{x},\vec{\sigma}_x^2,\vec{\gamma},\sigma_v^2,\beta_x) = p(\alpha_x|\vec{\sigma}_x^2,\beta_x)$. Analogous situations apply to the other steps.
\par
This sampling scheme can be viewed as a blocked Gibbs sampler, since the variables being sampled in steps 3, 4, and 5 are vector valued. Nevertheless, each variable is sampled exactly once using the corresponding posterior distribution conditioned on the current values of all other variables, hence, it is a valid Gibbs sampler. Due to our selection of conjugate priors, each conditional posterior distribution is analytically tractable. We now present the closed-form expressions for the sampling distributions in each step of Table \ref{table1}. The derivations are provided in the supplemental material.
\subsubsection{Sampling Distributions for Step 1 and 2}
The prior distribution parameters $\alpha_x$ and $\beta_x$ for the latent variable $\vec{\sigma}_x^2$ are sampled respectively from
\begin{equation}\label{sampling_alpha_x}
    p(\alpha_x|\vec{\sigma}_x^2,\beta_x) \propto \dfrac{\beta^{K\alpha_x}}{\Gamma(\alpha_x)^K}\bigg(\prod_{n=0}^{K-1}\dfrac{1}{\sigma_{x_n}^2}\bigg)^{\alpha_x+1}p(\alpha_x)
\end{equation}
and 
\begin{equation}\label{sampling_beta_x}
    p(\beta_x|\vec{\sigma}_x^2,\alpha_x) = \mathcal{G}(\beta_x;\tilde{\alpha}_{\beta_x},\tilde{\beta}_{\beta_x}),
\end{equation}
where $\tilde{\alpha}_{\beta_x} = K\alpha_x$ and $\tilde{\beta}_{\beta_x} = \sum_{n=0}^{K-1}1/\sigma_{x_n}^2$.
While sampling $\beta_x$ is straightforward, it is not for $\alpha_x$ due to not well-known form of its sampling distribution. Nevertheless, univariate form of (\ref{sampling_alpha_x}) allows us to draw samples efficiently by employing different sampling approaches, such as Metropolis-Hastings or Slice sampling \cite{RNeal}. 
\subsubsection{Sampling Distribution for Step 3} The posterior distribution of $\vec{\sigma}_x^2$ conditioned on $\vec{x}$, $\alpha_x$, and $\beta_x$ is given by
\begin{equation}\label{sampling_sigma_x}
    p(\vec{\sigma}_x^2|\vec{x},\alpha_x,\beta_x) = \prod_{n=0}^{K-1}\mathcal{IG}(\sigma_{x_n}^2;\tilde{\alpha}_x,\tilde{\beta}_{x_n})
\end{equation}
with common shape parameter $\tilde{\alpha}_x = \alpha_x + 1/2$ and individual scale parameters $\tilde{\beta}_{x_n} = x_n^2/2 + \beta_x$. Therefore, sampling $\vec{\sigma}_x^2$ can be achieved by independently sampling its elements from univariate IG distributions.
\subsubsection{Sampling Distribution for Step 4} The sampling distribution for the sparse sequence $\vec{x}$ takes the form of a multivariate Gaussian distribution
\begin{equation}\label{sampling_x}
    p(\vec{x}|\vec{y},\vec{\sigma}_x^2,\vec{\gamma},\sigma_v^2) = \mathcal{N}(\vec{x};\vec{\tilde{\mu}}_x,\vec{\tilde{\Sigma}}_x)
\end{equation}
with the posterior mean $\vec{\tilde{\mu}}_x$ and covariance $\vec{\tilde{\Sigma}}_x$ given by
\begin{equation}\label{sampling_x_mean_and_covariance}
    \vec{\tilde{\mu}}_x = \dfrac{1}{\sigma_v^2}\vec{\tilde{\Sigma}}_x\vec{H}^T\vec{y}, \qquad
    \vec{\tilde{\Sigma}}_x = \bigg(\dfrac{1}{\sigma_v^2}\vec{H}^T\vec{H} + \vec{\Sigma}_x^{-1}\bigg)^{-1}.
\end{equation}
Hence, sampling $\vec{x}$ is straightforward. 
\subsubsection{Sampling Distribution for Step 5} Similar to that of $\vec{x}$, the sampling distribution for $\vec{\gamma}$ is also a multivariate Gaussian distribution
\begin{equation}\label{sampling_gamma}
    p(\vec{\gamma}|\vec{y},\vec{x},\sigma_v^2) = \mathcal{N}(\vec{\gamma};\vec{\tilde{\mu}}_{\gamma},\vec{\tilde{\Sigma}}_{\gamma}),
\end{equation}
where the posterior mean and covariance are now given by
\begin{equation}\label{sampling_gamma_mean_and_covariance}
    \vec{\tilde{\mu}}_{\gamma} = \dfrac{1}{\sigma_v^2}\vec{\tilde{\Sigma}}_{\gamma}\vec{B}^T\vec{y},\qquad
    \vec{\tilde{\Sigma}}_{\gamma} = \bigg(\dfrac{1}{\sigma_v^2}\vec{B}^T\vec{B} + \dfrac{1}{\sigma_{\gamma}^2}\vec{I}\bigg)^{-1},
\end{equation}
with $\vec{B} = \vec{X}\vec{A}$, from which sampling is straightforward.
\subsubsection{Sampling Distribution for Step 6} The sampling distribution for the noise variance $\sigma_v^2$ takes the form of IG distribution given by
\begin{equation}\label{sampling_sigma_v}
    p(\sigma_{v}^2|\vec{y},\vec{x},\vec{\gamma}) = \mathcal{IG}(\sigma_v^2;\tilde{\alpha}_{v},\tilde{\beta}_v) 
\end{equation}
with parameters $\tilde{\alpha}_v = N/2$ and $\tilde{\beta}_v =  \|\vec{y}-\vec{X}\vec{A}\vec{\gamma}\|^2/2$. \par 

The use of alternative \text{soft} NIG prior for sparsity, instead of the BG model, allows us to introduce Step 3, where the latent variable $\vec{\sigma}_x^2$ is sampled conditioned on the current value of $\vec{x}$. The equivalent step for the BG model would be sampling the binary sequence $\vec{s}$ from $p(\vec{s}|\vec{x}) = p(\vec{x}|\vec{s})p(\vec{s})$, which would, however, not possible due to the deterministic dependence between $\vec{s}$ and $\vec{x}$. Therefore, $\vec{s}$ and $\vec{x}$ are usually sampled jointly. However, since $\vec{s}$ is a binary sequence of length $K$, sampling $\vec{s}$ as a whole at one step, which would require $2^K$ different probability mass point calculation, is not feasible. This enforces BG models to sample a single tuple $(s_n,x_n)$ at a time conditioned on the others. Since neighboring variables usually have strong dependence, sampling them conditioned on each other causes the sampler to be stuck on a local optimum for a long time. There is an extension proposed in \cite{DGe} that samples $M$-tuples at a time to escape from local optimums faster, but it is limited to very short blocks due to exponentially increasing computational load \cite{DGe}. Therefore, the discrete nature of the BG model creates essential computational burdens. By employing the alternative NIG model, the problem is transformed into a fully continuous valued framework, which not only helps to eliminate the exponential computational complexity but also provides an easier transition between different local optimums. 

\subsection{The Proposed Partially Collapsed Gibbs Sampler}
There exist two main points worth addressing in an attempt to improve the convergence rate of the proposed classical Gibbs sampler. Firstly, note that Step 3 samples $\vec{\sigma}_x^2$ purely conditioned on $\vec{x}$, which may slow down the convergence rate due to their strong statistical dependence. We address this issue by introducing intermediate proposal steps after Step 3, in which new values for coinciding blocks of $\vec{x}$ and $\vec{\sigma}_x^2$ are proposed using their joint posterior densities. Secondly, the blind nature of the problem creates many distinct local optimums corresponding to different $\vec{x}$ and $\vec{\gamma}$ pairs. Once a suboptimal\footnote{Here, by suboptimality, we refer to the configurations of $\vec{x}$ and $\vec{\gamma}$ that achieves a likelihood value as high as the true values of $\vec{x}$ and $\vec{\gamma}$.} configuration has been reached with a corresponding noise level $\sigma_v^2$, the sampler can get stuck on this configuration for a long time since $\vec{x}$, $\vec{\gamma}$, and $\sigma_v^2$ are all sampled conditioned on each other. This problem is especially important when the noise variance $\sigma_v^2$ gets smaller because it results in a posterior distribution with sharper and more isolated peaks. One way to resolve this issue is to sample $\vec{x}$ and $\vec{\gamma}$ jointly, which can be realized by marginalizing either $\vec{x}$ or $\vec{\gamma}$ from the sampling distributions given in Step 6 or 4, respectively. However, both of these approaches lead to complicated sampling distributions, from which sampling is not feasible due to high dimensionality. A less effective but more feasible way is to sample either $\vec{x}$ and $\sigma_v^2$, or $\vec{\gamma}$ and $\sigma_v^2$ jointly, creating more freedom for sampling $\sigma_v^2$. This allows $\sigma_v^2$ to assume larger values more frequently. Note that as $\sigma_v^2$ gets larger, the effect of likelihood is reduced on the conditional posteriors of both $\vec{x}$ or $\vec{\gamma}$, and it becomes easier to escape from a local optimum. Since $L \ll K$, we choose to sample $\vec{\gamma}$ and $\sigma_v^2$ jointly, because it is computationally much more efficient compared to sampling $\vec{x}$ and $\sigma_v^2$ jointly.\par
\begin{table}
\centering
\normalsize
\caption{Partially Collapsed Gibbs Sampler}
\vspace{-2mm}
\label{table2}
\renewcommand\arraystretch{1.3}
\begin{tabular}{|m{0.46\textwidth}|}
\hline
Step 1. Sample $\alpha_x$ from $p(\alpha_x|\vec{\sigma}_x^2,\beta_x)$  \\
Step 2. Sample $\beta_x$ from $p(\beta_x|\vec{\sigma}_x^2,\alpha_x)$  \\
Step 3. Sample $\vec{\sigma}_{x}^2$ from $p(\vec{\sigma}_{x}^2|\vec{x},\alpha_x,\beta_x)$
 \\
Step 4. Sample $\vec{x}_{\sim\ell_n}$ from $p(\vec{x}_{\sim\ell_n}|\vec{y},\vec{\sigma}_{x}^2,\vec{\gamma},\sigma_v^2)$   \\
Step 5. Sample $\vec{\sigma}_{x_{\ell_n}}^2$ from $p(\vec{\sigma}_{x_{\ell_n}}^2|\vec{y},\vec{x}_{\sim\ell_n},\vec{\gamma},\sigma_v^2,\alpha_x,\beta_x)$  \\
Step 6. Sample $\vec{x}_{\ell_n}$ from $p(\vec{x}_{\ell_n}|\vec{y},\vec{x}_{\sim\ell_n},\vec{\sigma}_{x_{\ell_n}}^2,\vec{\gamma},\sigma_v^2)$  \\
Step 7. Sample $\sigma_{v}^2$ from $p(\sigma_{v}^2|\vec{y},\vec{x})$  \\
Step 8. Sample $\vec{\gamma}$ from $p(\vec{\gamma}|\vec{y},\vec{x},\sigma_v^2)$  \\ \hline
\end{tabular}
\vspace{-5mm}
\end{table}
Let us first introduce the intermediate sampling steps for blocks of $\vec{x}$ and $\vec{\sigma}_x^2$. We define $\ell_n$ as the right-hand neighborhood of length $Q$ for index $n$, i.e., $\ell_n = \{n,n+1,\hdots,n+Q-1\}$ for $n = 0,1,\hdots,K-Q$, and let ${\sim}\ell_n$ be the complement of $\ell_n$ in $\{0,\hdots,Q-1\}$. We represent the variable blocks pointed by the neighborhood $\ell_n$ as $\vec{\sigma}_{x_{\ell_n}}^2 = [\sigma_{x_n}^2,\sigma_{x_{n+1}}^2,\hdots,\sigma_{x_{n+K-1}}^2]^T$ and $\vec{x}_{\ell_n} = [x_n,x_{n+1},\hdots,x_{n+K-1}]^T$. Similarly, $\vec{\sigma}_{x_{\sim\ell_n}}^2$ and $\vec{x}_{\sim\ell_n}$ represent the blocks of the remaining $K-Q$ variables. At each iteration of the new sampler, we propose new values for the blocks $\vec{\sigma}_{x_{\ell_n}}^2$ and $\vec{x}_{\ell_n}$ using the proposal distribution $p(\vec{\sigma}_{x_{\ell_n}}^2,\vec{x}_{\ell_n}|\vec{y},\vec{x}_{\sim\ell_n},\vec{\sigma}_{x_{\sim\ell_n}}^2,\vec{\gamma},\sigma_v^2)$, which is the joint posterior distribution of $\vec{\sigma}_{x_{\ell_n}}^2$ and $\vec{x}_{\ell_n}$ conditioned on all the other variables. Since this is actually a valid Gibbs sampling step, the Metropolis-Hastings acceptance probability is always 1 for the proposals. The neighborhood $\ell_n$ is updated after each iteration as follows. At $i^{th}$ iteration we set $n = \text{mod}(i-1,K-Q+1)$, where $\text{mod}(a,b)$ is the modulo operator returning the remainder after division of $a$ by $b$. This creates a sliding window over $\vec{\sigma}_x^2$ and $\vec{x}$ that shifts one index to the right at each iteration. Hence, the whole sequence is scanned after every $K-Q+1$ iterations.\par 
For a given neighborhood $\ell_n$, sampling from the joint conditional posterior $p(\vec{\sigma}_{x_{\ell_n}}^2,\vec{x}_{\ell_n}|\vec{y},\vec{x}_{\sim\ell_n},\vec{\gamma},\sigma_v^2,\alpha_x,\beta_x)$ can be realized in two consecutive steps, i.e., 
\begin{itemize}[noitemsep,topsep=1pt]
    \setlength\itemsep{0.1em}
    \item First, sample $\vec{\sigma}_{x_{\ell_n}}^2$ from $p(\vec{\sigma}_{x_{\ell_n}}^2|\vec{y},\vec{x}_{\sim\ell_n},\vec{\gamma},\sigma_v^2,\alpha_x,\beta_x)$,
    \item Then, sample $\vec{x}_{\ell_n}$ from $p(\vec{x}_{\ell_n}|\vec{y},\vec{x}_{\sim\ell_n},\vec{\sigma}_{x_{\ell_n}}^2,\vec{\gamma},\sigma_v^2)$.
\end{itemize}
These steps can be inserted right after Step 4, as Steps 5 and 6, respectively. Note that the block $\vec{x}_{\ell_n}$ is still being sampled in Step 4, which is redundant because it would not be conditioned on in Step 5 and immediately replaced with the new values obtained in Step 6. Therefore, sampling $\vec{x}_{\ell_n}$ can be skipped in Step 4, which forms a new step where only $\vec{x}_{\sim\ell_n}$ is sampled from $p(\vec{x}_{\sim\ell_n}|\vec{y},\vec{\sigma}_x^2,\vec{\gamma},\sigma_v^2)$. Although $\vec{\sigma}_{x_{\ell_n}}^2$ is also re-sampled in Step 5, it is still conditioned on in Step 4, so we cannot skip sampling $\vec{\sigma}_{x_{\ell_n}}^2$ in Step 3. The first 6 steps of the resulting sampling scheme is given in Table \ref{table2}.\par

In order to jointly sample $\sigma_v^2$ and $\vec{\gamma}$, we need to sample from $p(\sigma_v^2,\vec{\gamma}|\vec{y},\vec{x})$, which can also be realized in two steps as:
\begin{itemize}[noitemsep,topsep=1pt]
    \setlength\itemsep{0.1em}
    \item First, sample $\sigma_v^2$ from $p(\sigma_v^2|\vec{y},\vec{x})$,
    \item Then, sample $\vec{\gamma}$ from $p(\vec{\gamma}|\vec{y},\vec{x},\sigma_v^2)$.
\end{itemize}
These constitute the last two steps of the proposed sampler in Table \ref{table2}. \par 
We emphasize that the sampling distribution in Step 4 is not associated with the joint posterior $p(\vec{\theta}|\vec{y})$ anymore. Therefore, the proposed sampler is not a classical Gibbs sampler. Instead, it is a PCG sampler since the procedure described above is completely consistent with the \textit{marginalization} and \textit{trimming} operations described in \cite{DDyk}. PCG samplers are generalizations of the block Gibbs sampler, where some of the variables are not sampled from their full conditional posteriors. This provides more freedom for the sampler to jump from one point to another in the sampling space, which usually increases the mixing rate.\par 
Compared to the classical Gibbs sampler in Table \ref{table1}, the first three steps are the same and sampling distribution for Step 8 is already given in (\ref{sampling_gamma}). The sampling distributions for the remaining steps are presented below. Derivations are provided in the supplemental material. 
\subsubsection{Sampling Distribution for Step 4}\label{PCG_step_4}
The sampling distribution for $\vec{x}_{\sim \ell_n}$ block is obtained by marginalizing $\vec{x}_{\ell_n}$ from $p(\vec{x}|\vec{y},\vec{\sigma}_x^2,\vec{\gamma},\sigma_v^2)$. The resulting distribution is also a multivariate Gaussian distribution given by
\begin{equation}\label{sampling_x_block}
   p(\vec{x}_{\sim\ell_n}|\vec{y},\vec{\sigma}_x^2,\vec{\gamma},\sigma_v^2) = \mathcal{N}(\vec{x}_{\sim\ell_n};\vec{\tilde{\mu}}_{\sim\ell_n},\vec{\tilde{\Sigma}}_{\sim\ell_n}).
\end{equation}
The posterior mean $\vec{\tilde{\mu}}_{\sim\ell_n}$ and covariance $\vec{\tilde{\Sigma}}_{\sim\ell_n}$ are defined as
\begin{subequations}
\begin{align}
\vec{\tilde{\mu}}_{\sim\ell_n} &= \dfrac{1}{\sigma_v^2}\vec{\tilde{\Sigma}}_{\sim\ell_n}\vec{H}_{\sim\ell_n}^T\vec{D}_{\ell_n}^T\vec{y},\\
\vec{\tilde{\Sigma}}_{\sim\ell_n} &= \bigg(\dfrac{1}{\sigma_v^2}\vec{H}_{\sim\ell_n}^T\vec{D}_{\ell_n}\vec{H}_{\sim\ell_n} + \vec{\Sigma}_{\sim\ell_n}^{-1}\bigg)^{-1},
\end{align}
\end{subequations}
where $\vec{D}_{\ell_n} = \vec{I} - \frac{1}{\sigma_v^2}\vec{H}_{\ell_n}\vec{\tilde{\Sigma}}_{\ell_n}\vec{H}_{\ell_n}^T$ with $\vec{H}_{\sim\ell_n}$ and $\vec{H}_{\ell_n}$ represent the matrices formed by the columns of $\vec{H}$ indexed by ${\sim}\ell_n$ and $\ell_n$ respectively. The covariance matrices are defined as $\vec{\Sigma}_{\sim\ell_n} = \text{diag}(\vec{\sigma}_{x_{\sim\ell_n}}^2)$, $\vec{\Sigma}_{\ell_n} = \text{diag}(\vec{\sigma}_{x_{\ell_n}}^2)$, and $\vec{\tilde{\Sigma}}_{\ell_n} = (\frac{1}{\sigma_v^2}\vec{H}_{\ell_n}^T\vec{H}_{\ell_n} + \vec{\Sigma}_{\ell_n}^{-1})^{-1}$.

\subsubsection{Sampling Distribution for Step 5}\label{PCG_step_5}
Similarly, the sampling distribution for $\vec{\sigma}_{x_{\ell n}}^2$ block is obtained by marginalizing $\vec{x}_{\ell_n}$ from $p(\vec{\sigma}_{x_{\ell_n}}^2,\vec{x}_{\ell_n}|\vec{y},\vec{x}_{\sim\ell_n},\vec{\gamma},\sigma_v^2,\alpha_x,\beta_x)$, which is given by 
\begin{equation}\label{sampling_sigma_x_block}
\begin{split}
p(\vec{\sigma}_{x_{\ell_n}}^2|\vec{y},&\vec{x}_{\sim\ell_n},\vec{\gamma},\sigma_v^2,\alpha_x,\beta_x) \\
&\propto \dfrac{|\vec{\tilde{\Sigma}}_{\ell_n}|^{1/2}}{|\vec{\Sigma}_{\ell_n}|^{1/2}}\exp\bigg(\dfrac{1}{2}\vec{\tilde{\mu}}_{\ell_n}^T\vec{\tilde{\Sigma}}_{\ell_n}^{-1}\vec{\tilde{\mu}}_{\ell_n}\bigg)p(\vec{\sigma}_{x_{\ell_n}}^2).
\end{split}
\end{equation}
where $\vec{\tilde{\mu}}_{\ell_n} = \frac{1}{\sigma_v^2}\vec{\tilde{\Sigma}}_{\ell_n}\vec{H}_{\ell_n}^T\vec{\tilde{y}}_{\sim\ell_n}$ and $\vec{\tilde{y}}_{\sim\ell_n} = \vec{y} - \vec{H}_{\sim\ell_n}\vec{x}_{\sim\ell_n}$. Direct sampling from (\ref{sampling_sigma_x_block}) is not possible, since the form of the distribution is not well-known. Therefore, we employ a MH or Slice sampling step.

\subsubsection{Sampling Distribution for Step 6}
The sampling distribution for $\vec{x}_{\ell_n}$ block is given by
\begin{equation}\label{sampling_x_block_2}
    p(\vec{x}_{\ell_n}|\vec{y},\vec{x}_{\sim\ell_n},\vec{\sigma}_{x_{\ell_n}}^2,\vec{\gamma},\sigma_v^2) = \mathcal{N}(\vec{x}_{\ell_n};\vec{\tilde{\mu}}_{\ell_n},\vec{\tilde{\Sigma}}_{\ell_n}),
\end{equation}
with $\vec{\tilde{\mu}}_{\ell_n}$ and $\vec{\tilde{\Sigma}}_{\ell_n}$ are as defined in Section \ref{PCG_step_4} and \ref{PCG_step_5}.\par 

\subsubsection{Sampling Distribution for Step 7} The sampling distribution for the noise variance $\sigma_v^2$ is obtained by marginalizing $\vec{\gamma}$ from $p(\sigma_v^2,\vec{\gamma}|\vec{y},\vec{x})$, which is given by
\begin{align}\label{sampling_sigma_v_2}
    &p(\sigma_v^2|\vec{y},\vec{x}) \notag \\
    &\propto \bigg(\dfrac{1}{\sigma_v^2}\bigg)^{N/2}|\vec{\tilde{\Sigma}}_{\gamma}|^{1/2}\exp\bigg(\dfrac{1}{2}\vec{\tilde{\mu}}_{\gamma}^T\vec{\tilde{\Sigma}}_{\gamma}^{-1}\vec{\tilde{\mu}}_{\gamma} - \dfrac{1}{2\sigma_v^2} \vec{y}^T\vec{y}\bigg)p(\sigma_v^2).
\end{align}
Although it is not straightforward to sample from (\ref{sampling_sigma_v_2}), we can efficiently employ MH or Slice sampling methods due to its univariate form. In the next section, we investigate the scaling and time-shifting ambiguities existing in the BD problems and propose two intermediate sampling steps accounting these ambiguities.

\subsection{Scale and Time-Shift Ambiguities}
Scaling and time-shift ambiguities are inherent in blind deconvolution problems preventing unique recovery of the pulse shape and the sparse sequence. For a given solution pair $(\vec{x},\vec{h})$, one can produce infinitely many different solutions consisting of the scaled versions $(\alpha\vec{x},\vec{h}/\alpha)$ with $\alpha \in \mathbbm{R}$, which constitutes the scale ambiguity. The time-shifted versions $(\vec{x} \ast \delta_n,\vec{h} \ast \delta_{-n})$, where $\delta_n$ is the Kronecker delta with spike at position $n$ and $\ast$ denotes the convolution operation, is another source of ambiguity constituting the time-shift ambiguity. In practice, recovery of the true parameters up to an arbitrary scale and time-shift does not cause a major problem and is usually sufficient. 
\subsubsection{Scale Ambiguity}
In Bayesian framework, assignment of the prior distributions can eliminate the scaling ambiguity only if the scaling of the prior distributions is fixed at an anchor point. Otherwise, with such a scaling transformation $\{a\vec{x},a^2\vec{\sigma}_x^2,\alpha_x,a^2\beta_x,\vec{\gamma}/a,\sigma_{\gamma}^2/a^2,\sigma_v^2\}$ with $a > 0$ that leaves the likelihood invariant, one can scale the posterior with $a^{L-K+4}$. This means that the posterior distribution can be increased arbitrarily by decreasing the scale $a$, indicating nonexistence of a global optimum. However, this issue can be avoided by setting, for example, $\sigma_{\gamma}^2$ to a fixed constant. In this case, the sampler eventually converges to a fixed scale $\alpha'$ associated with the value of $\sigma_{\gamma}^2$. However, convergence might be slow since the parameters are sampled conditioned on each other. A common approach to accelerate the convergence is to introduce an intermediate Metropolis-Hastings (MH) sampling step. Following this approach, once the current values of $(\vec{x},\vec{\gamma},\vec{\sigma}_x^2)$ are sampled, we first propose the new values for $\vec{x}^{*}$ and $\vec{\gamma}^{*}$, i.e., $(\vec{x}^{*},\vec{\gamma}^{*}) = (\alpha\vec{x},\vec{\gamma}/\alpha)$, by sampling the scaling factor $\alpha$ from the proposal distribution $q(\alpha) = \mathcal{N}(\alpha;0,\sigma_{\alpha}^2)$ with known variance $\sigma_{\alpha}^2$ and then propose $\vec{\sigma}_x^{2*}$ by sampling from $p(\vec{\sigma}_x^{2*}|\vec{x}^*,\alpha_x,\beta_x)$, yielding the complete proposal distribution $q(\vec{x}^{*},\vec{\gamma}^{*},\vec{\sigma}_x^{2*}|\vec{x},\vec{\gamma},\vec{\sigma}_x^2) = q(\alpha)p(\vec{\sigma}_x^{2*}|\vec{x}^*,\alpha_x,\beta_x)$. The proposed values are accepted with probability $p_{\alpha}$, which is given by
\begin{equation}\label{MH_scaling_acceptance_prob}
p_{\alpha} = \min\bigg\{1,\alpha^L\exp\bigg(\dfrac{\alpha^{4}-1}{2\alpha^{2}\sigma_{\alpha}^2}\bigg)\prod_{n=0}^{K-1}\bigg[\dfrac{x_n^2 + 2\beta_x}{\alpha^2x_n^2 + 2\beta_x}\bigg]^{\alpha_x + 0.5} \bigg\}
\end{equation}
We note that a better strategy would be adjusting the parameters $\alpha_x$ and $\beta_x$ as well by sampling new values $\alpha_x^{*}$ and $\beta_x^{*}$ from $p(\alpha_x^{*},\beta_x^{*}|\vec{\sigma}_x^{2*})$. However, achieving a closed form expression for the acceptance probability would not be possible.

\subsubsection{Time-Shift Ambiguity}
Unlike scaling ambiguity, time-shift ambiguity does not fully apply to our case due to the edge effects, i.e., the edges of the reconstructed observation sequence will be corrupted when shifted versions of a given solution are considered. This is because we model the observation sequence using linear convolution of finite length sequences. Therefore, solution pairs with different shifts cannot achieve exactly the same likelihood value. However, although large time-shift ambiguities are explicitly avoided, time-shifts of very short lengths can still correspond to a similar level of likelihood, and hence, needs to be addressed. Since both sequences $\vec{x}$ and $\vec{h}$ are sampled conditioned on the current value of the other, jumps between different time-shift configurations rarely occur. To increase the frequency of these jumps, we employ the circular shift compensation method proposed in \cite{CLabat}. Following the scale compensation step, new values of $\vec{x}$, $\vec{\sigma}_x^2$ and $\vec{\gamma}$ are proposed using the proposal distribution $ q(\vec{x}^{*},\vec{\sigma}_x^{2*},\vec{\gamma}^{*}|\vec{x},\vec{\sigma}_x^{2},\vec{\gamma}) = q(\vec{x}^{*},\vec{\sigma}_x^{2*}|\vec{x},\vec{\sigma}_x^2)p(\vec{\gamma}^{*}|\vec{y},\vec{x}^{*},\sigma_v^2)$
where
\begin{equation}\label{MH_time_shift_proposal_2}
    q(\vec{x}^{*},\vec{\sigma}_x^{2*}|\vec{x},\vec{\sigma}_x^2) = 
    \begin{cases}
        0.5 \; \text{if} \; (\vec{x}^*,\vec{\sigma}_x^{2*}) = (\vec{x}\circledast\delta_{-1},\vec{\sigma}_x^{2}\circledast\delta_{-1})\\
        0.5 \; \text{if} \; (\vec{x}^*,\vec{\sigma}_x^{2*}) = (\vec{x}\circledast\delta_{1},\vec{\sigma}_x^{2}\circledast\delta_{1})
    \end{cases}
\end{equation}
and $\circledast$ denotes the circular convolution. It can be shown that the MH acceptance probability $p_s$ is given by
\begin{equation}\label{MH_time_shift_acceptance_prob}
p_{s} = \min\bigg\{1,\dfrac{|\vec{\tilde{\Sigma}}_{\gamma^*}|}{|\vec{\tilde{\Sigma}}_{\gamma}|}\exp\bigg(\dfrac{1}{2}\vec{\tilde{\mu}}_{\gamma^*}^T\vec{\tilde{\Sigma}}_{\gamma^*}^{-1}\vec{\tilde{\mu}}_{\gamma^*} - \dfrac{1}{2}\vec{\tilde{\mu}}_{\gamma}^T\vec{\tilde{\Sigma}}_{\gamma}^{-1}\vec{\tilde{\mu}}_{\gamma}\bigg)\bigg\}
\end{equation}
where both $(\vec{\tilde{\mu}}_{\gamma^*},\vec{\tilde{\Sigma}}_{\gamma^*})$ and $(\vec{\tilde{\mu}}_{\gamma},\vec{\tilde{\Sigma}}_{\gamma})$ can be calculated through (\ref{sampling_gamma_mean_and_covariance}) using $\vec{x}^*$ and $\vec{x}$ respectively.

\section{Simulations}\label{simulations}
In this section, we present our numerical studies to assess the performance of the proposed samplers and compare our results with the classical BG deconvolution approach, in which the sparsity of $\vec{x}$ is enforced by the BG prior, as described in (\ref{conditional_prior_x_given_s})-(\ref{joint_prior_x_and_s}), instead of the NIG prior used in this work. The form of the slightly modified version of Cheng et al.'s classical Gibbs sampler employing BG model is given in Table \ref{table3}. Step 1 of this sampler consists of $K$ sub-update steps, where pairs of $(s_n,x_n)$ are sampled jointly from their joint posterior distribution at each sub-step. Jointly sampling $(s_n,x_n)$ is also completed in two steps, first sampling $s_n$ from $p(s_n|\vec{y},\vec{x}_{\sim n},\vec{\gamma},\sigma_v^2)$ and then sampling $x_n$ from $p(x_n|\vec{y},\vec{x}_{\sim n},s_n,\vec{\gamma},\sigma_v^2)$. The remaining sampling steps are the same as the proposed samplers. We also consider the $M$-tuple Gibbs sampler proposed in \cite{DGe}, with $M = 3$. It modifies Step 1 of Table \ref{table3} to sample tuples of length $3$ at a time and  considerably improves the convergence rate of classical BGS. Hence, it constitutes a stronger baseline for assessing the performance of our samplers. Throughout the simulations, we use these as our baseline samplers and call them, respectively, as the classical Bernoulli-Gaussian Sampler (BGS) and $3$-Tuple BGS. We refer to the proposed samplers using the Normal-Inverse-Gamma law as NIGS-1 (for the classical Gibbs sampler) and NIGS-2 (for the partially collapsed Gibbs sampler). \par
\begin{table}
\centering
\normalsize
\caption{Baseline Classical Bernoulli-Gaussian Sampler}
\vspace{-2mm}
\label{table3}
\renewcommand\arraystretch{1.2}
\begin{tabular}{|p{0.06\textwidth} p{0.35\textwidth}|}
\hline
Step 1. & For $n =0,1,\hdots,K-1$, sample $(s_n,x_n)$ from $p(s_n,x_n|\vec{y},\vec{x}_{\sim n},\vec{\gamma},\sigma_v^2)$\\
Step 2. & Sample $\vec{\gamma}$ from $p(\vec{\gamma}|\vec{y},\vec{x},\sigma_v^2)$  \\
Step 3. & Sample $\sigma_v^2$ from $p(\sigma_v^2|\vec{y},\vec{x},\vec{\gamma})$   \\ \hline
\end{tabular}
\vspace{-5mm}
\end{table}
Due to the scale and minor time-shift ambiguities that are inherent in blind deconvolution problems, all recovery results presented in this section are up to an arbitrary scale and shift factor. The compensation steps introduced in the previous section resolve these ambiguities only for the sampling stage, i.e., the resulting estimated sequences are not necessarily expected to match the scale and shift of the true sequences. Therefore, for any given estimation pair $(\hat{\vec{x}},\hat{\vec{h}})$, we present the corrected versions $\vec{x}^{\prime}$ and $\vec{h}^{\prime}$, which are given by $\vec{x}^{\prime} = (\hat{\vec{x}} \ast \delta_{-n})/a$ and $\vec{h}^{\prime} = a(\hat{\vec{h}} \ast \delta_{n})$, where the correction factors are found using
\begin{equation}\label{scale_and_time_shift_correction}
(a,n) = \argmin_{a\in \mathbbm{R},n\in \mathbb{Z}}\|\vec{h} - a(\hat{\vec{h}} \ast \delta_n)\|.
\end{equation}

\begin{figure*}[t!]
  \begin{subfigure}[b]{0.33\textwidth}
    \includegraphics[width=0.95\textwidth]{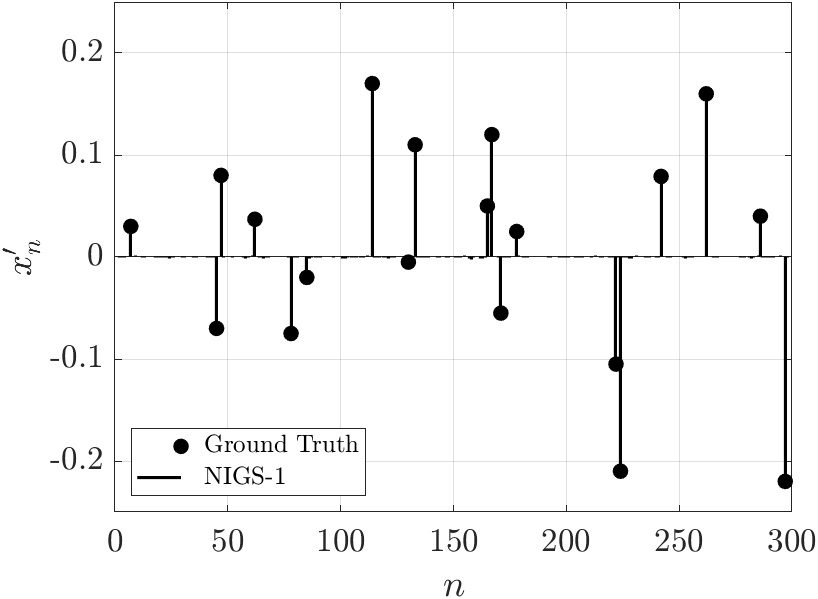}
  \end{subfigure}
  \begin{subfigure}[b]{0.33\textwidth}
    \includegraphics[width=0.95\textwidth]{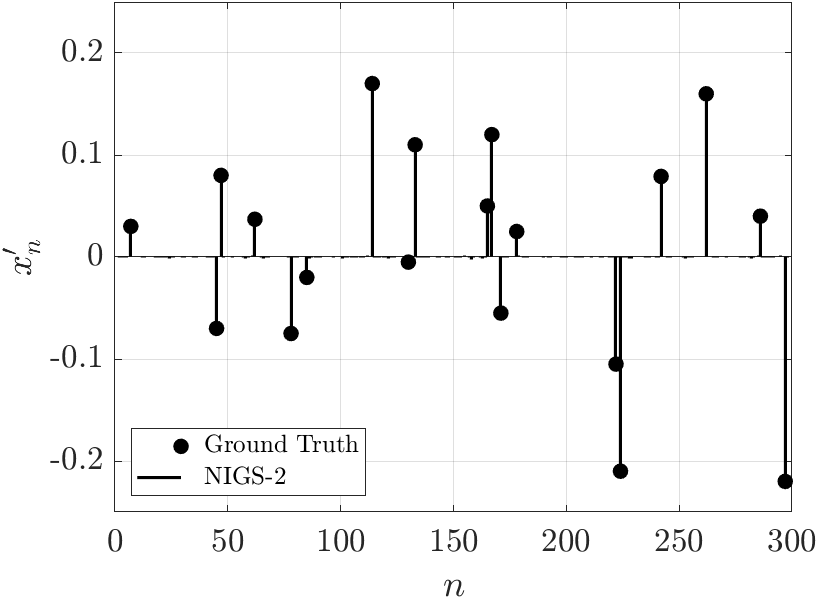}
  \end{subfigure}
    \begin{subfigure}[b]{0.33\textwidth}
    \includegraphics[width=0.95\textwidth]{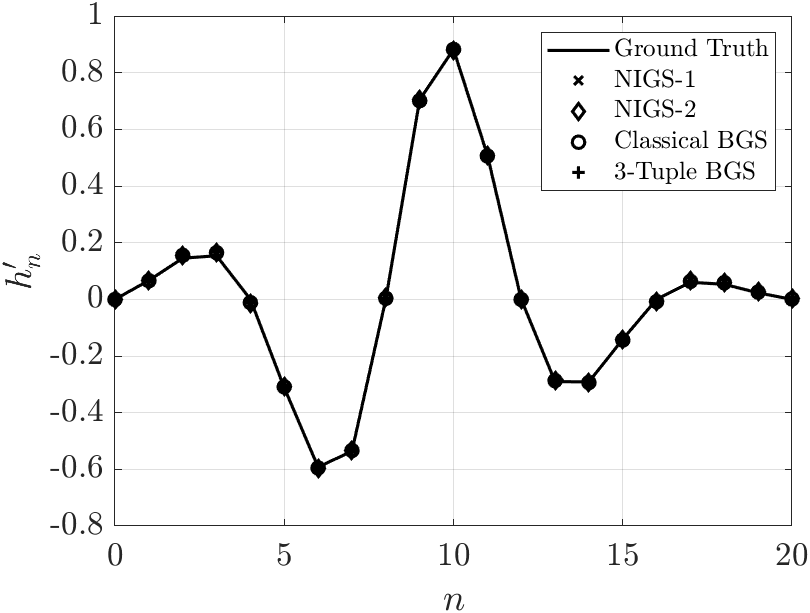}
  \end{subfigure}
    \begin{subfigure}[b]{0.33\textwidth}
    \includegraphics[width=0.95\textwidth]{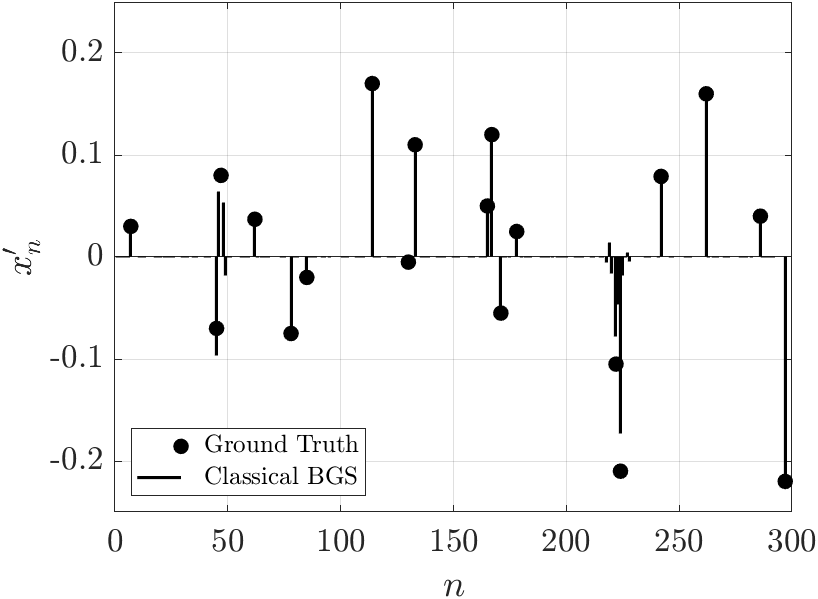}
  \end{subfigure}
    \begin{subfigure}[b]{0.33\textwidth}
    \includegraphics[width=0.95\textwidth]{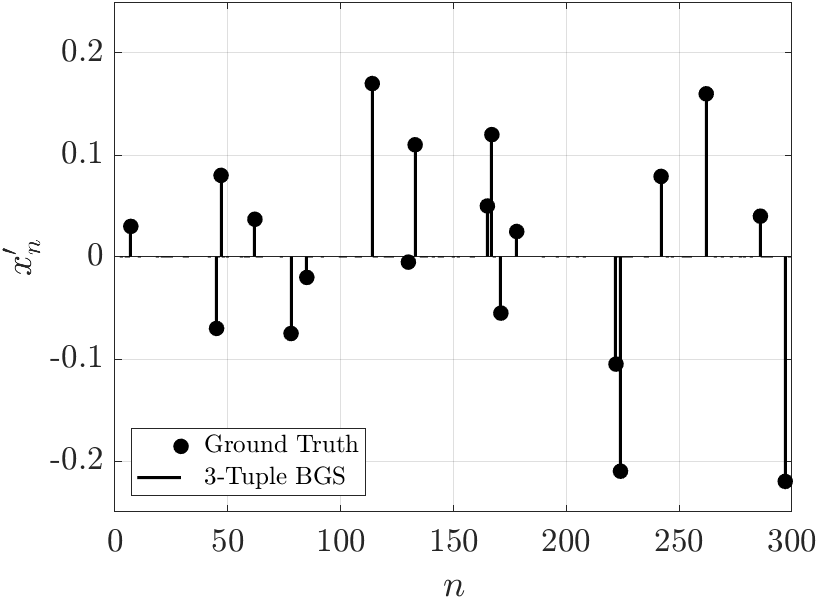}
  \end{subfigure}
    \begin{subfigure}[b]{0.33\textwidth}
    \includegraphics[width=0.95\textwidth]{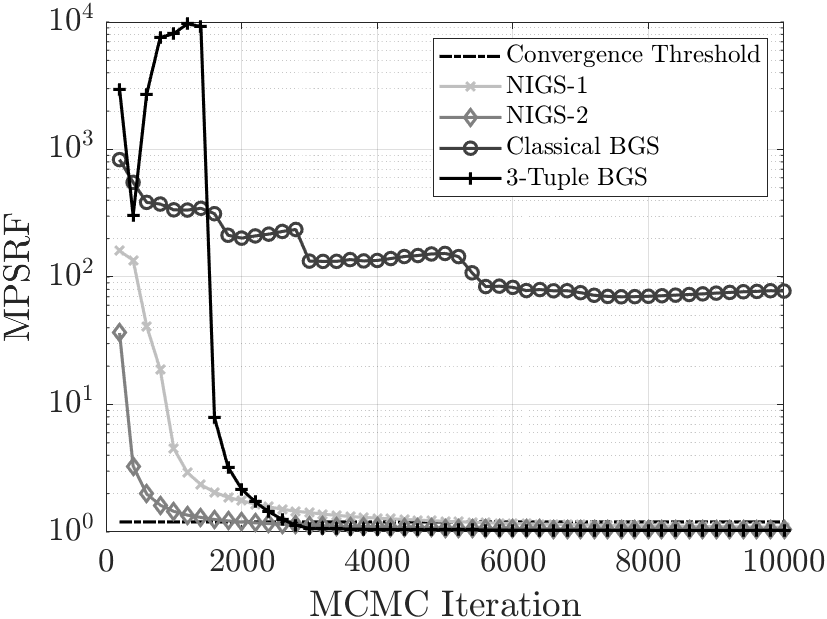}
  \end{subfigure}
  \caption{Recovery results for the sparse sequence (left and middle columns), pulse sequence (upper right), and convergence rates (lower right). \label{mendel_results}}
  \vspace{-9mm}
\end{figure*}

\subsection{A Measure of Convergence Rate}
In order to empirically measure the convergence rate, we employ the iterated graphical monitoring approach proposed by Brook and Gelman in \cite{SBrooks}. The diagnostic is based on comparing the between and within chain covariance matrices, denoted by $\vec{S}_b$ and $\vec{S}_w$ respectively, of different simulation chains that are running simultaneously. More precisely, for multivariate simulations, convergence is declared when the multivariate potential scale reduction factor (MPSRF), denoted by $\hat{R}$, becomes close to 1. A typical threshold in which the MPSRF is expected to fall below is 1.2 as suggested in \cite{SBrooks}. The MPSRF is defined as 
\begin{equation}\label{MPSRF}
    \hat{R} = \dfrac{i-1}{i} + \dfrac{q+1}{q}\lambda_{\text{max}},
\end{equation}
where $i$ denotes the number of MCMC iterations, $q$ is the total number of distinct simulation chains running simultaneously and $\lambda_{\text{max}}$ is the maximum eigenvalue of the matrix $\vec{S}_w^{-1}\vec{S}_b$. The definitions of the between chain covariance matrix $\vec{S}_b$ and the within chain covariance matrix $\vec{S}_w$ are given as
\begin{subequations}
\begin{align}
    \vec{S}_b &= \dfrac{1}{q-1}\sum_{j=1}^{q}(\vec{\bar{\theta}}_{.j} - \vec{\bar{\theta}}_{..})(\vec{\bar{\theta}}_{.j} - \vec{\bar{\theta}}_{..})^T,\\
    \vec{S}_w &= \dfrac{1}{q(i-1)}\sum_{j=1}^{q}\sum_{l=1}^{i}(\vec{\theta}_{lj} - \vec{\bar{\theta}}_{.j})(\vec{\theta}_{lj} - \vec{\bar{\theta}}_{.j})^T,
\end{align}
\end{subequations}
where $\vec{\theta}_{lj}$ denotes the $l^{th}$ sample of $j^{th}$ chain. Also, $\vec{\bar{\theta}}_{.j}$ and $\vec{\bar{\theta}}_{..}$ represent the local mean of the $j^{th}$ chain and the global mean of the all chains respectively. Therefore, for a given data, we run several chains from different initial points and monitor the MPSRF value to assess the convergence.  

\subsection{Recovery Performance on Mendel Sequence}\label{results_mendel}
As an illustrative example, we present the recovery results of the proposed  samplers on a given observation sequence and provide a comparison with the classical BGS described above in order to show its inefficiency. The observation sequence $y_n$ was generated based on the linear convolution model given in (\ref{conv_model}). As the sparse sequence $x_n$, we used the well-known Mendel's sequence, which models a 1-D sparse reflectivity profile for seismic blind deconvolution \cite{JMendel}. The sequence is depicted with bullets in sparse recovery plots of Fig. \ref{mendel_results}. As the short pulse sequence $h_n$, we used the following sequence 
\begin{equation}\label{pulse_sequence}
    h_n = \cos\big((n-10)\pi/4\big)\exp\big(-|0.225n-2|^{1.5}\big)
\end{equation}
for $n = 0,1,\hdots,20$, which is the same sequence used in both \cite{QCheng} and \cite{DGe}. It is represented with solid line in the upper-right plot of Fig. \ref{mendel_results} with solid line. The noise variance was set as $\sigma_v^2 = 4.8\times10^{-6}$, which corresponds to a Signal-to-Noise Ratio (SNR) of 25 dB. 
\par
The lengths of the sparse sequence $\vec{x}$ and the pulse sequence $\vec{h}$ are $K = 300$ and $T = 21$, respectively, yielding a length $N = 321$ measurement sequence $\vec{y}$. For this experiment, we did not impose any frequency domain constraints on the pulse sequence and set the subspace matrix as identity, i.e., $\vec{A} = \vec{I}$. For Step 1 of NIGS-1, and Steps 1, 5, and 7 of NIGS-2, we employed univariate Slice sampling approach. We set the window size for NIGS-2 as $Q = 10$ and used the following fixed values for the parameters of baseline samplers: $\sigma_x^2 = 1$, $\sigma_{\gamma}^2 = 10$, and $\pi_0 = 1-|\vec{x}|_0/K$, where $|\vec{x}|_0$ denotes the number of nonzero elements in the true sequence $\vec{x}$. For all samplers, we generated $10^4$ samples and used the last $25\%$ for producing the estimations.
\par
The four plots in the left and middle columns of Fig. \ref{mendel_results} illustrate the sparse sequences recovered by each sampler. The recoveries obtained by the proposed samplers and 3-Tuple BGS are almost identical and perfectly match the true sequence. On the other hand, the sparse sequence recovered by the classical BGS contains inaccurate nonzero entries where the actual spikes are located closely. This example shows the main inefficiency of the classical BGS. Since the $(s_n,x_n)$ tuples are sampled conditioned on the current value of adjacent entries, it usually takes a large number of iterations to escape from a local optimum. Therefore, initialization plays an important role in the convergence behavior of BGS. This issue will become clearer once we present the convergence analysis.\par

The recovered pulses, as shown in middle lower plot in Fig. \ref{mendel_results}, are almost identical and perfectly match the true pulse shape for all samplers. This is usually the case since the number of parameters to be estimated, i.e., the degree of freedom, is much smaller for the pulse sequence.\par 

In order to compare the convergence rates of the samplers, we illustrate the evolution of the MPSRF corresponding to samples of $\vec{x}$ and $\vec{\gamma}$ for all samplers in the lower-right plot in Fig. \ref{mendel_results}. We simulated 10 independent chains with distinct initialization for each sampler and updated the MPSRF after every 200 iterations using the last 50\% of the generated samples. The MPSRF curves of the proposed samplers NIGS-1 and NIGS-2 consistently decrease and fall below the convergence threshold of $1.2$ after around $5000$ and $2000$ iterations, respectively. This suggests that each of the individual chains of NIGS-1 and NIGS-2 converged to the same local optimum, which is very likely to be the global one, regardless of where they are initialized at the parameter space.  However, the MPSRF curve of classical BGS fails to converge and cannot fall below the convergence threshold during the simulation duration. This result confirms that individual chains of BGS get stuck on a local optimum for a much larger number of iterations. Note that unlike classical BGS, 3-Tuple BGS reaches the convergence threshold at around $2500$ iterations, which is due to its improved convergence rate.

\begin{figure*}[t!]
\centering
  \begin{subfigure}[b]{0.48\textwidth}
  \centering
    \includegraphics[width=0.95\textwidth]{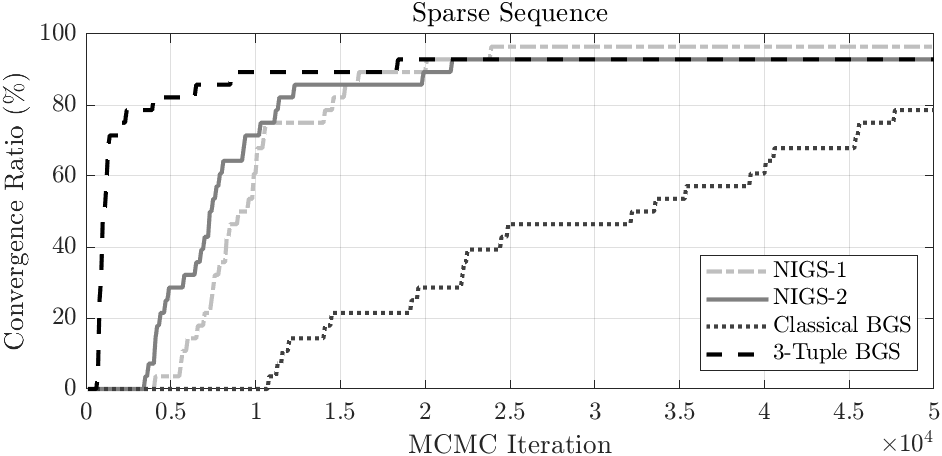}
  \end{subfigure}
  \begin{subfigure}[b]{0.48\textwidth}
  \centering
    \includegraphics[width=0.95\textwidth]{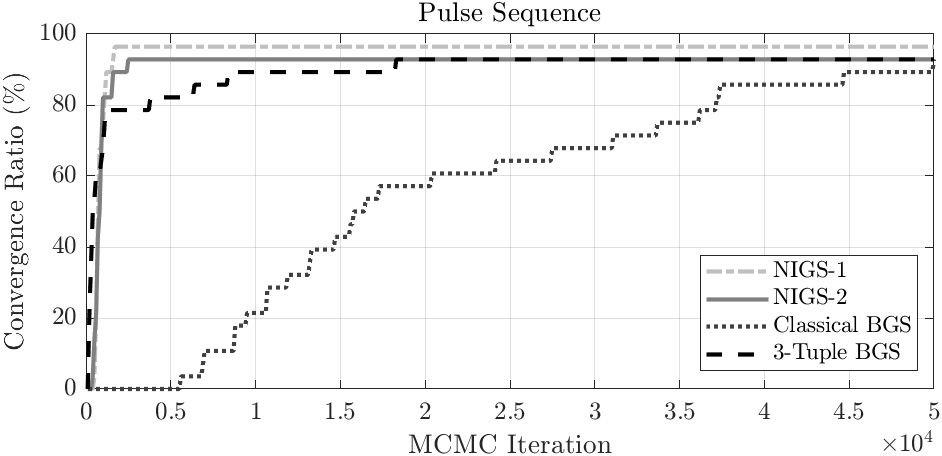}
  \end{subfigure}
  \begin{subfigure}[b]{0.48\textwidth}
  \centering
    \includegraphics[width=0.95\textwidth]{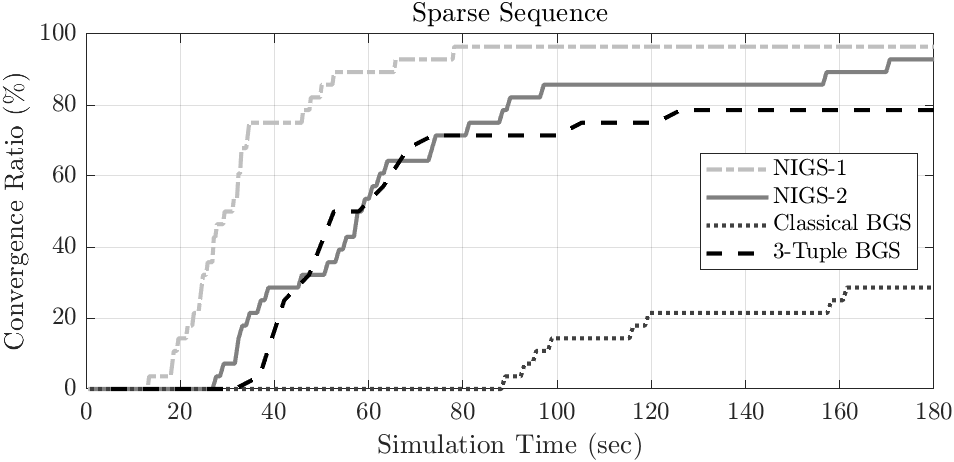}
  \end{subfigure}
  \begin{subfigure}[b]{0.48\textwidth}
  \centering
    \includegraphics[width=0.95\textwidth]{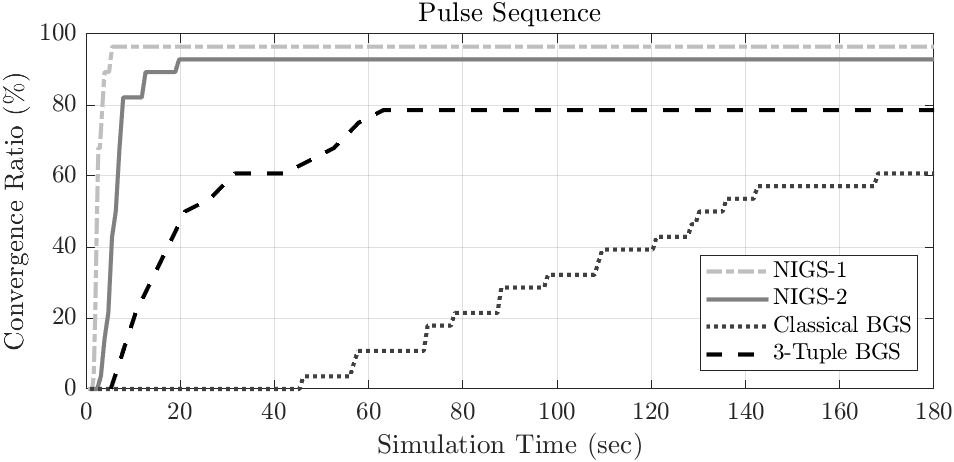}
  \end{subfigure}
  \caption{Ratio of the number of converged simulations to the total number of simulations as a function of MCMC iteration (top) and CPU time (bottom) for the sparse sequence $\vec{x}$ and the pulse sequence $\vec{h}$.\label{convergence_ratio}}
  \vspace{-5mm}
\end{figure*}

\subsection{Empirical Analysis of the Convergence Rates}\label{results_convergence_rates}
Next, we evaluate the convergence performance of the proposed samplers over a set of randomly generated sparse sequences. The generated sequences have length $K = 300$. The nonzero positions were randomly distributed across the sequence. In order to ensure that there are no localized regions with all nonzero elements, we divided each sequence into multiple segments of equal lengths and randomly choose a nonzero index for each segment. A $6\%$ sparsity level, defined by the ratio of the number of nonzero entries to the total length of the sequence, was maintained for all generated sequences. The amplitudes for the nonzero positions were independently drawn from a univariate zero-mean Gaussian distribution with a variance of 0.5. A total of 50 distinct observations were generated by convolving each of 50 different sparse sequences with the pulse sequence given in (\ref{pulse_sequence}), which were then corrupted by additive noise such that the SNR is 20 dB. 
\par
For each observation, we run all samplers 10 times with different initialization to obtain 10 independently simulated chains. The MPSRF values are calculated on these chains after every 100 MCMC iterations by using the last $50\%$ of the samples. Same as before, we set the convergence threshold for MPSRF as 1.2. In Fig. \ref{convergence_ratio}, we illustrate the ratio of converged sequences over all 50 sequences as a function of the MCMC iteration for all samplers. In order to individually investigate the convergence performance on different unknowns, we calculated the MPSRF curves separately for the sparse sequence $\vec{x}$ and pulse coefficients $\vec{\gamma}$, which are represented in the upper left and right figures, respectively. 
\par

We see in the top left figure that the ratio of converged sequences for the proposed samplers increases quite rapidly compared to classical BGS, while all three are outperformed by 3-Tuple BGS. The outstanding convergence performance of 3-Tuple BGS is mainly due to the fact that many of the configurations leading to a local optimum occur within a short neighborhood, which can be escaped rapidly even with a very small tuple size $M$. However, the computational complexity of $M$-Tuple BGS increases exponentially with the tuple size $M$ as shown in \cite{DGe}. We also observed that NIGS-2 achieves a slightly better convergence rate compared to NIGS-1, which indicates that the intermediate sampling steps introduced in NIGS-2 help accelerating the mixing rate. Comparing the left and right figures, we see that convergence rate is faster for all samplers in the case of pulse coefficients $\vec{\gamma}$. This is indeed an expected result since the degree of freedom is considerably higher for the sparse sequence. We further note that the convergence rates of the proposed samplers are almost as fast as that of 3-Tuple BGS, and they still significantly outperform classical BGS.

\subsection{Computational Complexity Analysis}
\begin{figure*}[t!]
  \begin{subfigure}[b]{0.48\textwidth}
  \centering
    \includegraphics[width=0.95\textwidth]{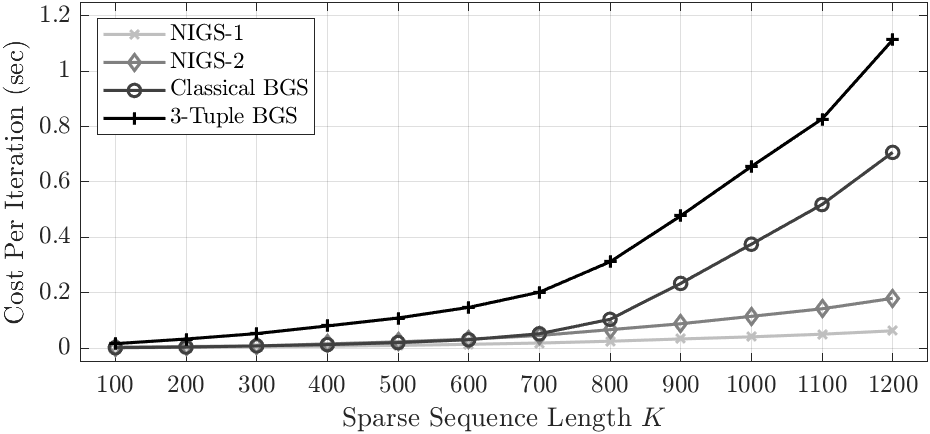}
  \end{subfigure}
  \begin{subfigure}[b]{0.48\textwidth}
  \centering
    \includegraphics[width=0.95\textwidth]{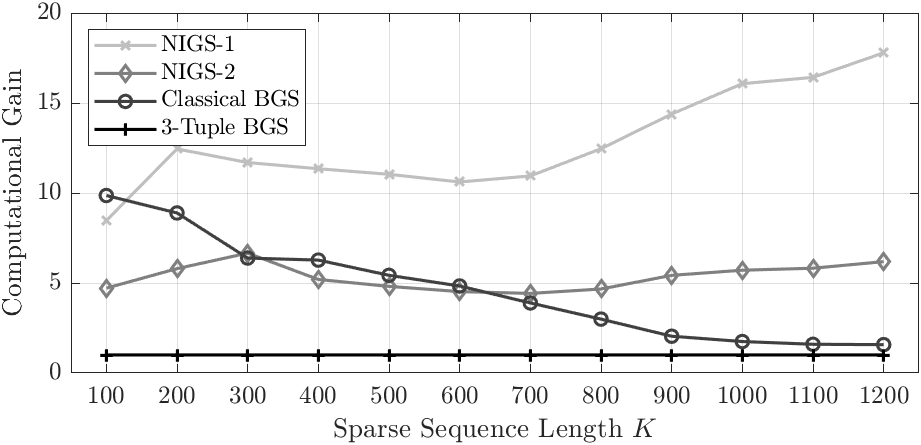}
  \end{subfigure}
  \caption{Comparison of cost per iteration in seconds for sparse sequences with varying lengths (left). Computational gain of the samplers relative to 3-Tuple BGS as a function of sparse sequence length $K$ (right). \label{computational_complexities}}
  \vspace{-1mm}
\end{figure*}

The study presented above compared the convergence rates of the different algorithms in terms of the number of MCMC iterations. Next, we incorporate the computational complexities of the samplers into the analysis. In order to investigate the empirical computational complexities, we measured the processing times based on unoptimized MATLAB R2019a implementation of the samplers on an Intel Core i5-8265U processor. For the implementation of classical and 3-Tuple BGS, we used the efficient numerical method presented in \cite{DGe}. The average elapsed time for one iteration of each sampler as a function of the sparse sequence length $K$ is illustrated in the left plot in Fig. \ref{computational_complexities}. For all considered values of $K$, the number of spikes was adjusted accordingly to maintain the 6\% sparsity level. It can be observed that the cost per iteration in terms of processing time is significantly higher for 3-Tuple BGS for all values of $K$. Moreover, the difference between the computational complexities of the proposed samplers and classical BGS increases considerably for larger values of $K$. In order to better illustrate the difference of computational complexities, we also present the computational gain of all samplers relative to 3-Tuple BGS in the right plot in Fig. \ref{computational_complexities}. We define the gain as the ratio of cost per iteration of 3-Tuple BGS to those of other samplers. It is clear that NIGS-1 and NIGS-2 are, respectively, at least around 10 and 5 times computationally more efficient compared to 3-Tuple BGS. Moreover, the computational gain of the proposed samplers increases for larger values of $K$. We also observed that classical BGS is computationally more efficient compared to NIGS-2 for smaller values of $K$, even though its cost per iteration increases significantly for larger values of $K$.\par

\begin{table}[t!]
\renewcommand\arraystretch{1.1}
  \centering
  \caption{Overall empirical successful recovery rates for the sparse sequence $\vec{x}$ (top) and the pulse sequence $\vec{h}$ (bottom). The NMSE threshold $\tau$ is varied between $0.01$ and $0.1$.\label{overall_success_rates}}
  \vspace{-2mm}
  \begin{tabularx}{\linewidth}{ C{0.2\linewidth} C{0.14\linewidth} C{0.14\linewidth} C{0.14\linewidth} C{0.14\linewidth} }
    \multicolumn{5}{c}{\small Sparse Sequence} \\
    \toprule
     & $\tau = 0.01$ & $\tau = 0.04$ & $\tau = 0.07$ & $\tau = 0.1$ \\
    \midrule
    \textbf{NIGS-1} & 0.54  & 0.69 & 0.75 & 0.79 \\ 
    \textbf{NIGS-2} & 0.57 & 0.72 & 0.78 & 0.81 \\
    \textbf{Classical BGS} & 0.36 & 0.53 & 0.62 & 0.69 \\
    \textbf{3-Tuple BGS} & 0.57 & 0.73 & 0.79 & 0.82 \\
    \bottomrule
  \end{tabularx}
  
  \medskip
  
  \begin{tabularx}{\linewidth}{ C{0.2\linewidth} C{0.14\linewidth} C{0.14\linewidth} C{0.14\linewidth} C{0.14\linewidth} }
    \multicolumn{5}{c}{\small Pulse Sequence} \\
    \toprule
     & $\tau = 0.01$ & $\tau = 0.04$ & $\tau = 0.07$ & $\tau = 0.1$ \\
    \midrule
    \textbf{NIGS-1} & 0.82  & 0.92 & 0.94 & 0.96 \\ 
    \textbf{NIGS-2} & 0.83  & 0.93 & 0.96 & 0.97 \\
    \textbf{Classical BGS} & 0.85 & 0.93 & 0.95 & 0.96 \\
    \textbf{3-Tuple BGS} & 0.86 & 0.93 & 0.95 & 0.96 \\
    \bottomrule
  \end{tabularx}
  \vspace{-5mm}
\end{table}

\begin{figure*}[t!]
\centering
  \begin{subfigure}[b]{0.24\textwidth}
    \includegraphics[width=\textwidth]{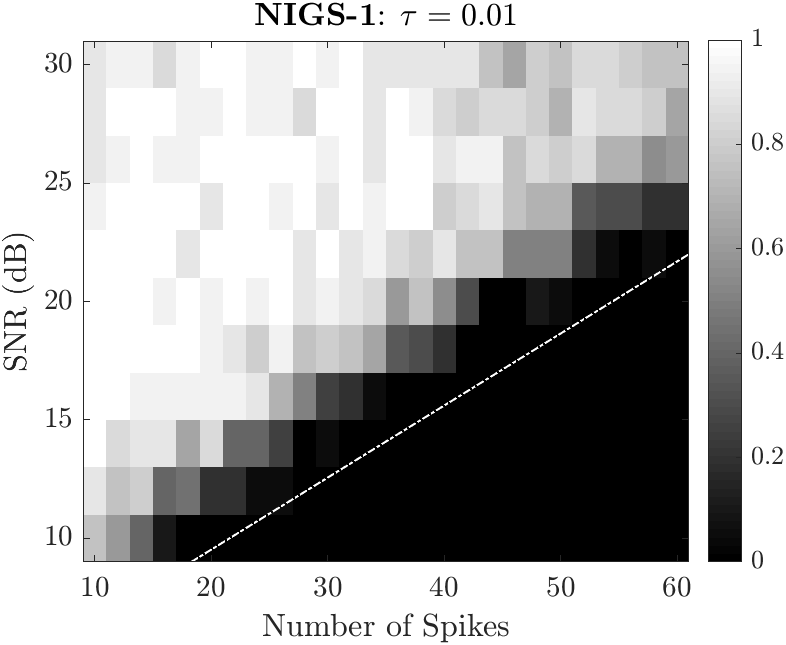}
  \end{subfigure}
    \begin{subfigure}[b]{0.24\textwidth}
    \includegraphics[width=\textwidth]{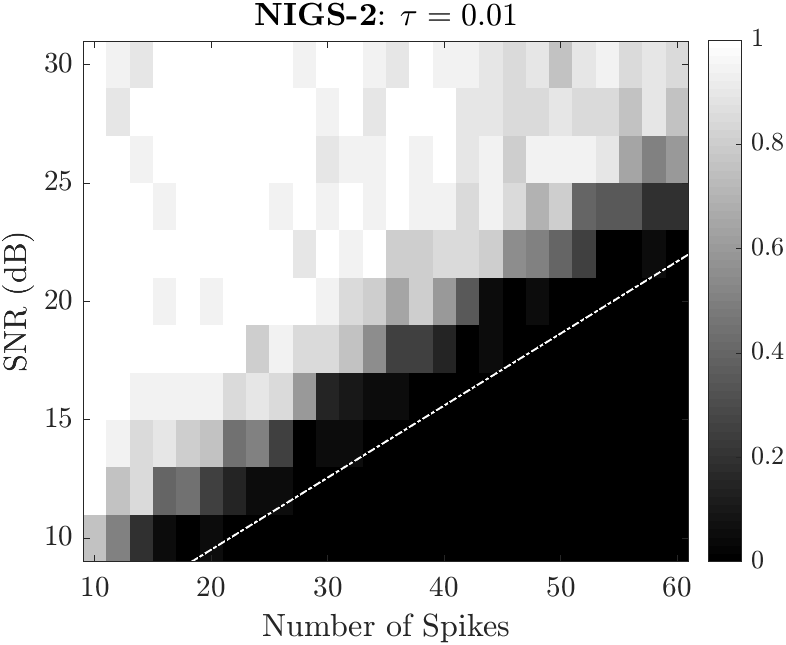}
  \end{subfigure}
  \begin{subfigure}[b]{0.24\textwidth}
    \includegraphics[width=\textwidth]{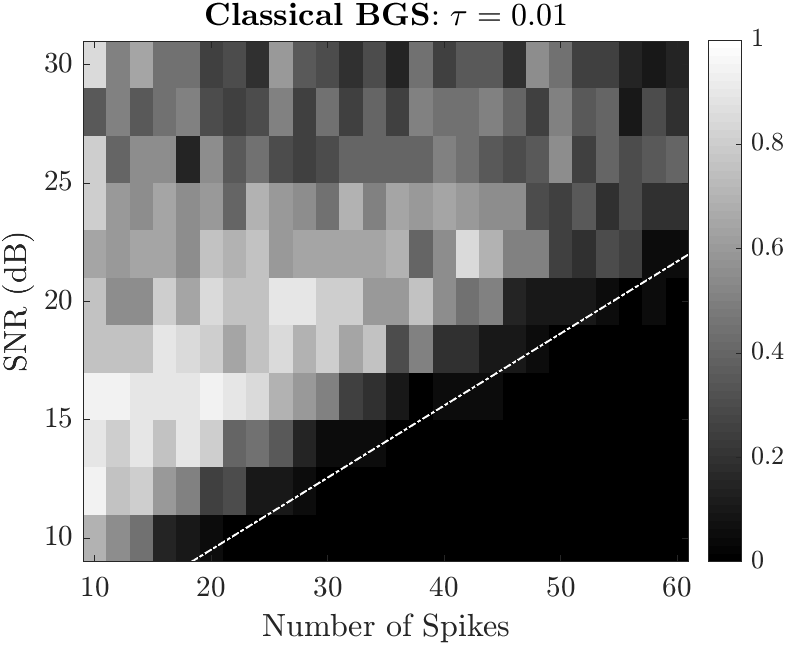}
  \end{subfigure}
    \begin{subfigure}[b]{0.24\textwidth}
    \includegraphics[width=\textwidth]{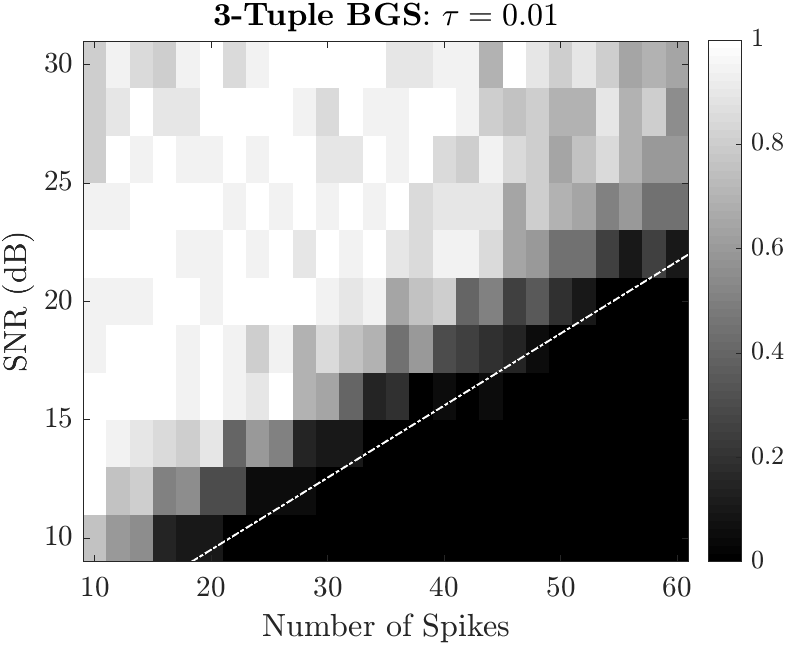}
  \end{subfigure}
    \begin{subfigure}[b]{0.24\textwidth}
    \includegraphics[width=\textwidth]{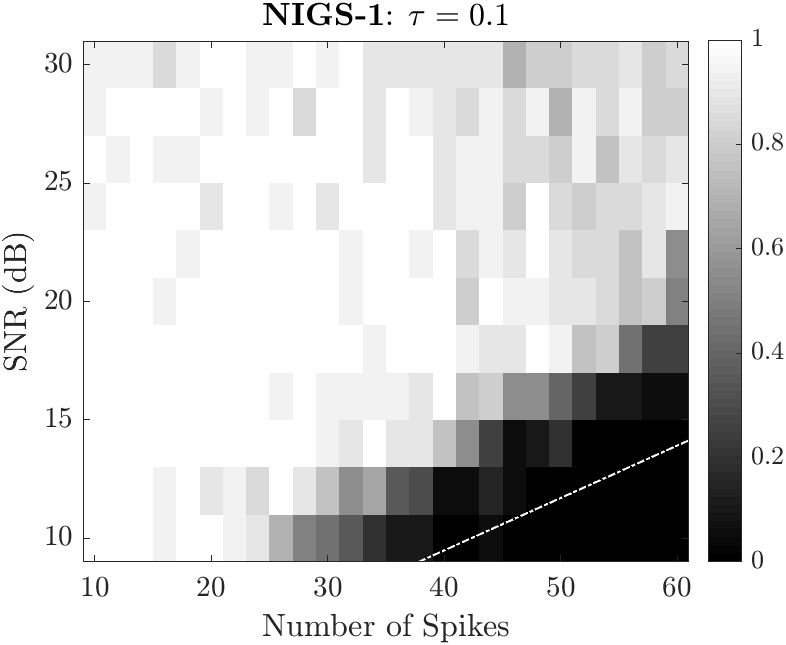}
  \end{subfigure}
    \begin{subfigure}[b]{0.24\textwidth}
    \includegraphics[width=\textwidth]{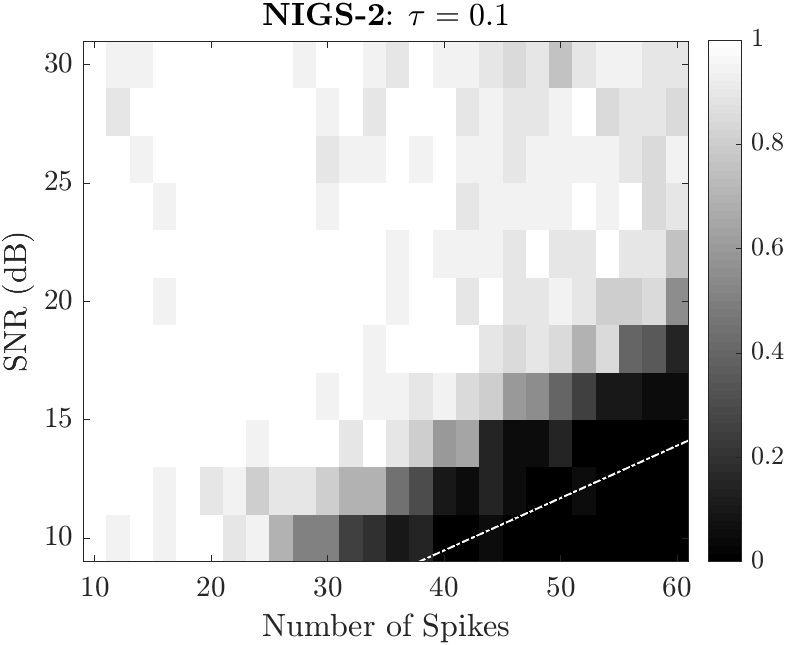}
  \end{subfigure}
    \begin{subfigure}[b]{0.24\textwidth}
    \includegraphics[width=\textwidth]{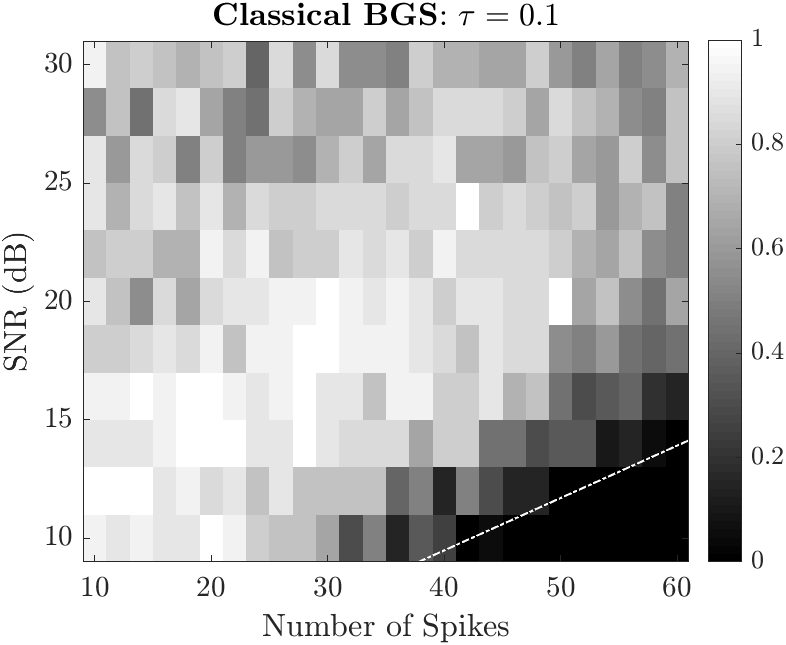}
  \end{subfigure}
  \begin{subfigure}[b]{0.24\textwidth}
    \includegraphics[width=\textwidth]{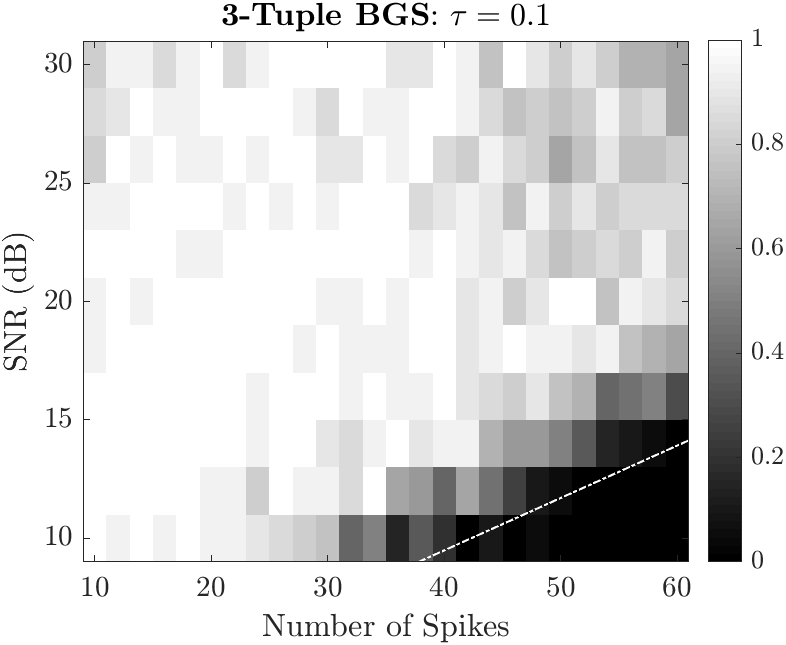}
  \end{subfigure}
  \caption{Empirical successful recovery rates for the sparse sequence $\vec{x}$ at different SNR and sparsity levels for $\tau = 0.01$ and $\tau = 0.1$.  Dashed white line depicts the transition boundary for 3-Tuple BGS. \label{success_rate_snr_vs_sparsity_x}}
  \vspace{-5mm}
\end{figure*}
Based on empirically measured computational costs of the samplers, we can also compare the convergence rates in terms of the simulation time. The comparisons are provided in bottom rows of Fig. \ref{convergence_ratio} for both sparse sequence $\vec{x}$ and the pulse coefficients $\vec{\gamma}$ using length $K = 300$ sparse sequences. The results indicate that the excessive computational cost of 3-Tuple BGS overwhelms its outstanding convergence rate. Due to its lower cost per iteration, NIGS-1 achieves the fastest convergence rate in terms of processing time for both $\vec{x}$ and $\vec{\gamma}$. NIGS-2 has a slightly higher cost per iteration compared to NIGS-1, but still attains a similar convergence rate to that of 3-Tuple BGS for $\vec{x}$, which is even better for $\vec{\gamma}$. Due to increasing computational gain with $K$, it would also be reasonable to suggest that the difference between convergence rates will be more substantial for larger values of $K$, favoring the use of proposed samplers in practice.

\subsection{Overall Recovery Performance for Different Scenarios}
In this section, we investigate the recovery performance of the proposed samplers is investigated under various different SNR and sparsity levels, along with the comparisons with the baseline samplers. We considered 11 different SNR levels ranging between 10 dB and 30 dB with 2 dB separation between each level. For each SNR level, we investigated 26 different scenarios, where the number of spikes is increased from 10 to 60 with 2 increments. The length of the sparse sequence for each scenario was fixed at $K=300$, yielding sparsity levels ranging from 3\% to 20\%. We created 20 different random sparse sequences for each one of these 286 scenarios using the same way described in Section \ref{results_convergence_rates}. For the pulse sequence, we used the normalized first derivative of Gaussian pulse, given by 
\begin{equation}
    h_n = 2(n\pi f_c/f_s - 2)\exp\big(0.5-2(n \pi f_c/f_s - 2)^2\big)
\end{equation}
for $n = 0,1,\hdots,22$ with center frequency $f_c = 2$ GHz and sampling rate $f_s = 36$ GHz. This pulse shape constitutes a strictly short duration sequence in the time domain, which is also nearly bandlimited in the frequency domain. Therefore, for these experiments, we construct the columns of the pulse subspace matrix $\vec{A}$ using the first $L = 8$ DPS sequences of length $T = 23$. We run each sampler for $10^4$ iterations and use the last 25\% of the samples for estimation of the unknowns. We declare a recovered sparse sequence $\vec{x}^{\prime}$ (or pulse sequence $\vec{h}^{\prime}$) successful if the Normalized Mean Squared Error (NMSE) is less than a given threshold $\tau$, i.e.,
\begin{equation}\label{success_condition}
    \dfrac{\|\vec{x} - \vec{x}^{\prime}\|^2}{\|\vec{x}\|^2} \leq \tau, \qquad \dfrac{\|\vec{h} - \vec{h}^{\prime}\|^2}{\|\vec{h}\|^2} \leq \tau.
\end{equation}

\begin{figure*}[t!]
\centering
  \begin{subfigure}[b]{0.24\textwidth}
    \includegraphics[width=\textwidth]{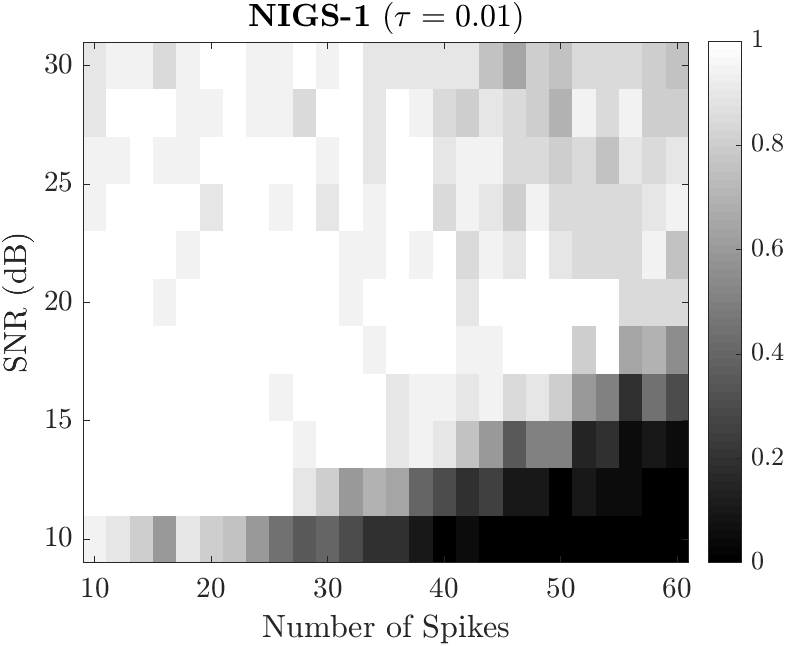}
  \end{subfigure}
    \begin{subfigure}[b]{0.24\textwidth}
    \includegraphics[width=\textwidth]{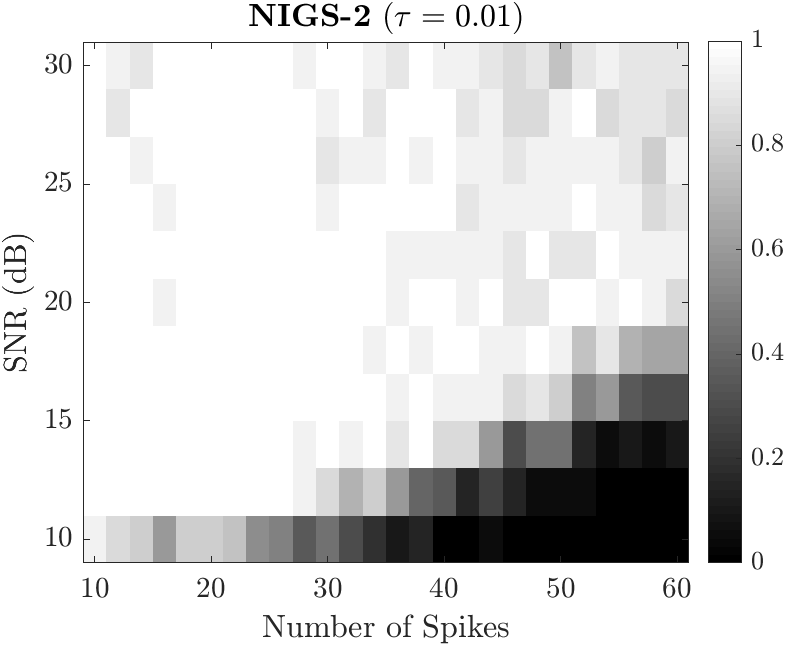}
  \end{subfigure}
  \begin{subfigure}[b]{0.24\textwidth}
    \includegraphics[width=\textwidth]{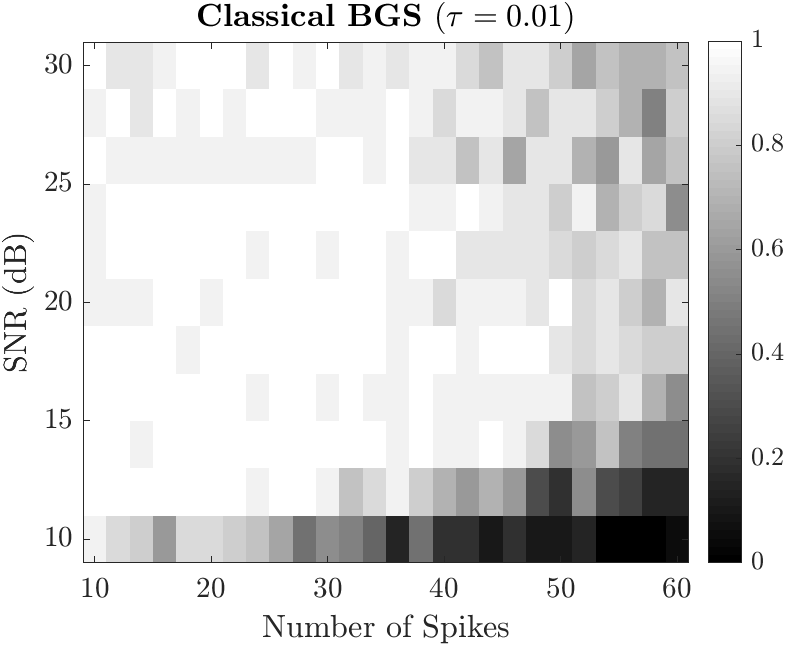}
  \end{subfigure}
    \begin{subfigure}[b]{0.24\textwidth}
    \includegraphics[width=\textwidth]{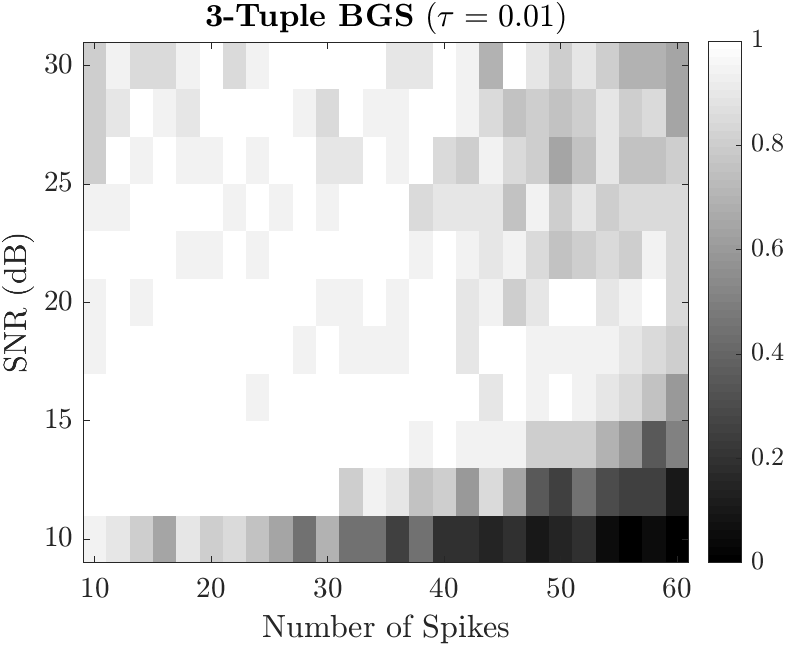}
  \end{subfigure}
    \begin{subfigure}[b]{0.24\textwidth}
    \includegraphics[width=\textwidth]{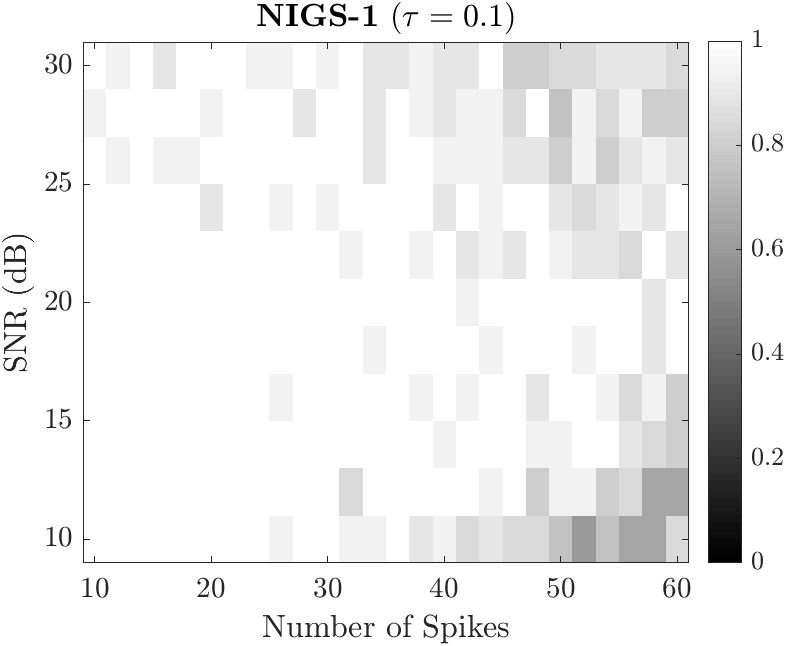}
  \end{subfigure}
    \begin{subfigure}[b]{0.24\textwidth}
    \includegraphics[width=\textwidth]{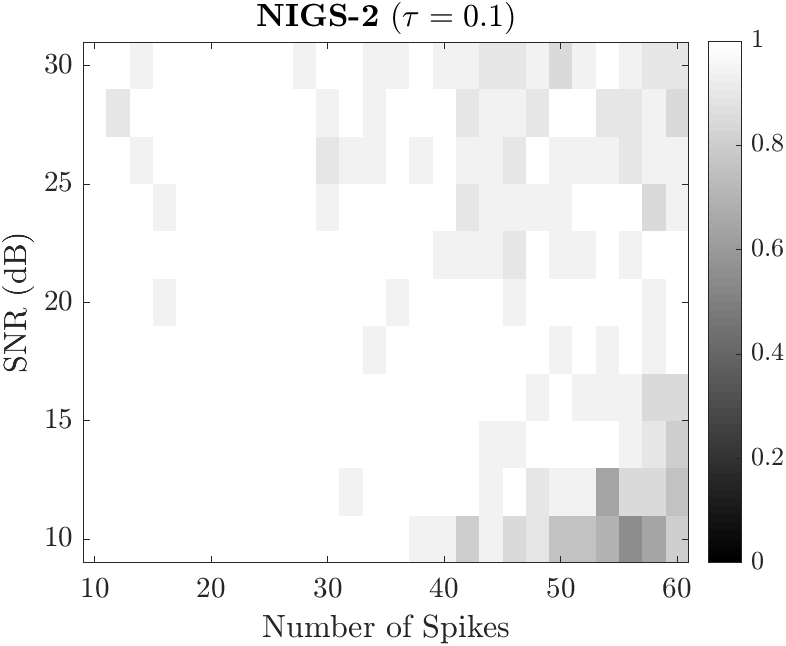}
  \end{subfigure}
    \begin{subfigure}[b]{0.24\textwidth}
    \includegraphics[width=\textwidth]{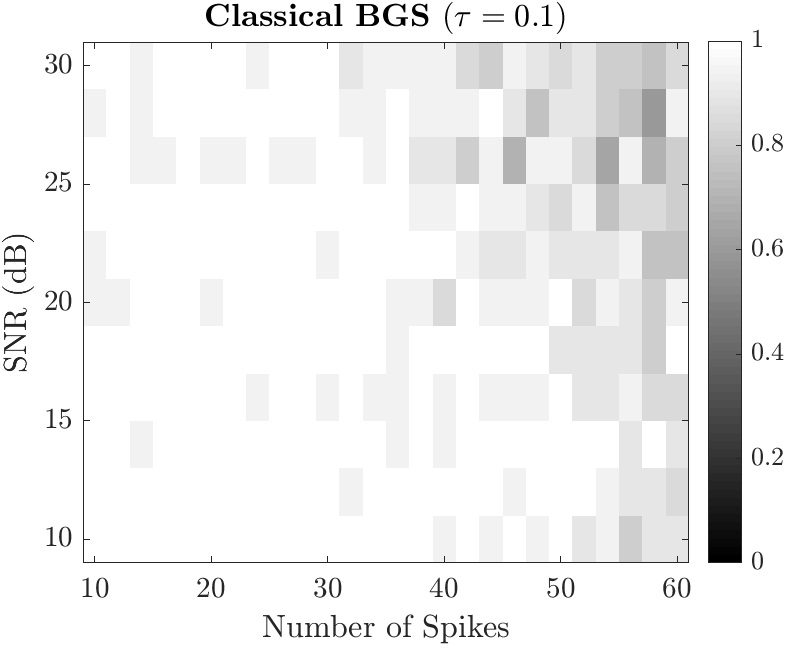}
  \end{subfigure}
  \begin{subfigure}[b]{0.24\textwidth}
    \includegraphics[width=\textwidth]{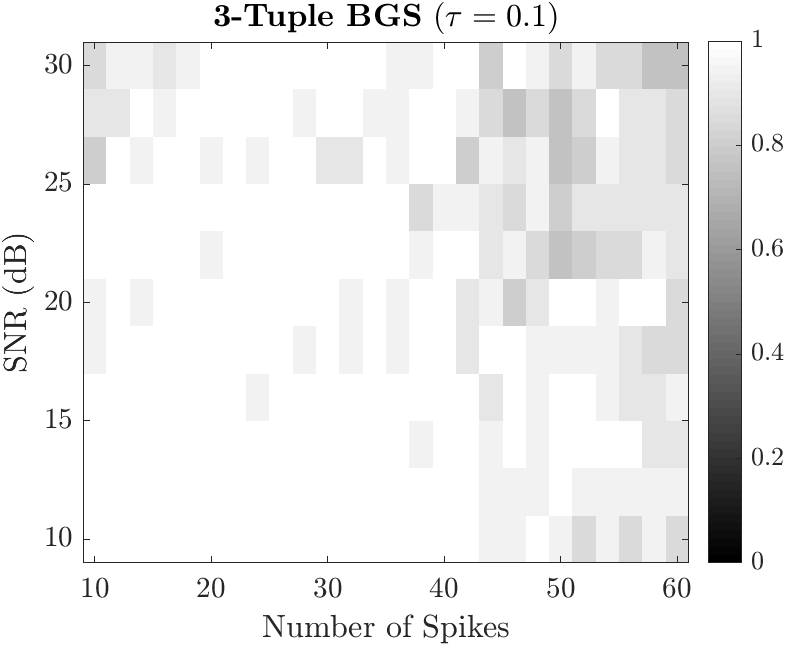}
  \end{subfigure}
  \caption{Empirical successful recovery rates for the pulse sequence $\vec{h}$ at different SNR and sparsity levels for $\tau = 0.01$ and $\tau = 0.1$. \label{success_rate_snr_vs_sparsity_h}}
  \vspace{-5mm}
\end{figure*}

\begin{figure*}[t!]
  \begin{subfigure}[b]{0.48\textwidth}
  \centering
    \includegraphics[width=0.9\textwidth]{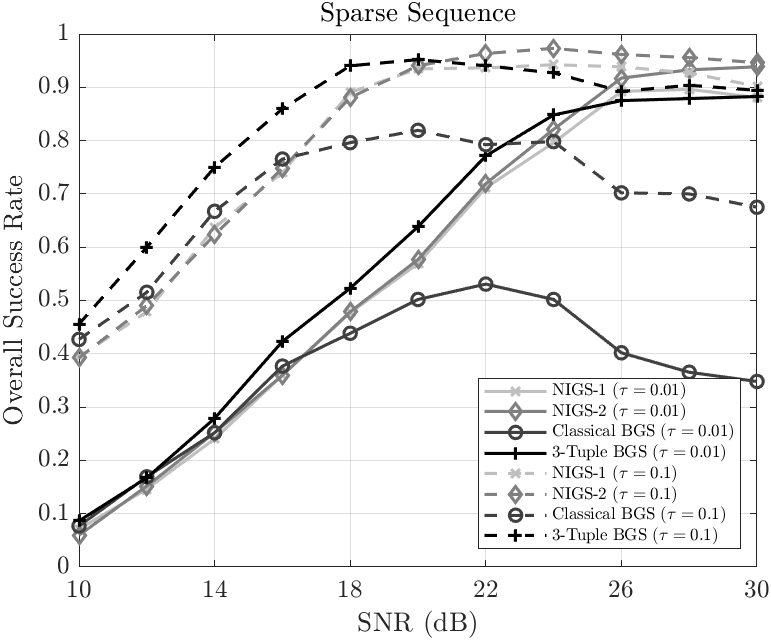}
  \end{subfigure}
  \begin{subfigure}[b]{0.48\textwidth}
  \centering
    \includegraphics[width=0.9\textwidth]{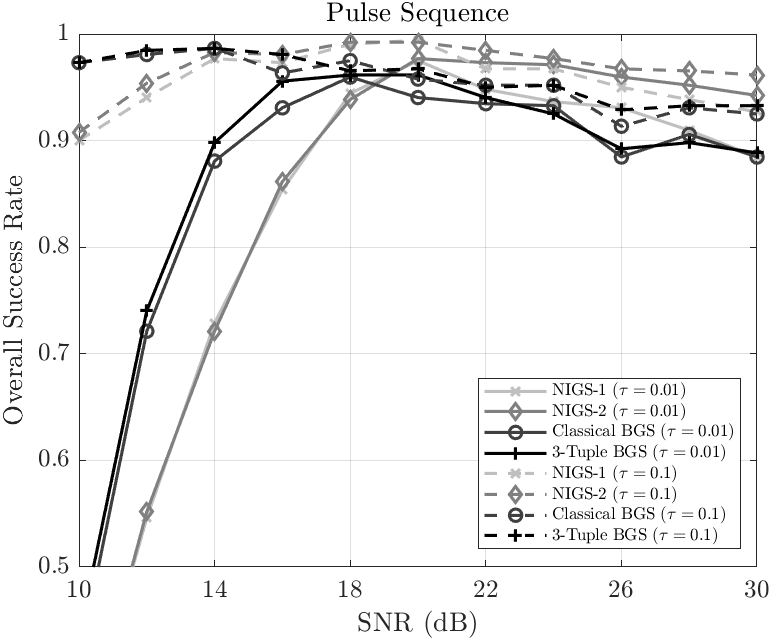}
  \end{subfigure}
  \caption{Overall empirical successful recovery rates for the sparse sequence $\vec{x}$ (left) and the pulse sequence $\vec{h}$ (right) at different SNR levels. Solid and dashed lines correspond to $\tau = 0.01$ and $\tau = 0.1$ cases respectively.\label{overall_success_rates_vs_SNR}}
  \vspace{-5mm}
\end{figure*}

In Fig. \ref{success_rate_snr_vs_sparsity_x}, we illustrate the empirical successful recovery rates of the samplers for the sparse sequence $\vec{x}$ for two different values of $\tau$. We first focus on the transition boundaries, which identify the feasible regions in the sparsity/SNR plane for which successful recovery, as defined in (\ref{success_condition}), is possible. In order to better illustrate the differences, we draw the transition boundary of 3-Tuple BGS, associated with the corresponding $\tau$, as a reference at all plots. It can be observed that feasible regions of the proposed samplers are slightly more restricted. This indicates that for a given sparsity level, the proposed samplers require a slightly higher SNR level for successful recovery. Nevertheless, the feasible regions are quite similar for all samplers. On the other hand, the rate of successful recovery within the feasible region is significantly higher for the proposed sampler as opposed to classical BGS. This is due to the fact that classical BGS requires a considerably larger number of iterations than $10^4$ to converge to the true stationary distribution. In addition, we also observed that NIGS-2 achieves the highest overall success rate within its feasible region even though 3-Tuple BGS has a slightly larger feasible region. This is also justified by the average success rates, given in Table \ref{overall_success_rates}, corresponding to different values of $\tau$, over all 286 scenarios. Despite the slightly smaller feasible region, NIGS-2 achieves a success rate as high as that of 3-Tuple BGS. The results also show that NIGS-2 achieves a higher overall success rate compared to NIGS-1, which provides empirical evidence that the intermediate sampling steps of NIGS-2 improves the convergence rate. Moreover, it is clear that both proposed samplers significantly outperform classical BGS. \par 

We present the corresponding success rates for the pulse shape $\vec{h}$ in Fig. \ref{success_rate_snr_vs_sparsity_h}. Comparing the transition plots in Fig. \ref{success_rate_snr_vs_sparsity_x} and \ref{success_rate_snr_vs_sparsity_h}, our first observation is that feasible regions for the pulse sequence are significantly more extensive for all samplers. As it can be seen from $\tau = 0.01$ case, similar to the recovery of sparse sequence, the feasible regions are more extensive for the baseline samplers. Another observation is that the success rates of classical BGS are significantly higher for the pulse sequence. This indicates that the main source of the inefficiency of classical BGS is different configurations of sparse sequences rather than the pulse shape. Table \ref{overall_success_rates} shows that all samplers achieve a similar level of success rate for the pulse sequence except for $\tau = 0.01$, which can be explained by the smaller feasible regions of the proposed samplers. Overall, the success rates are consistently higher compared to the recovery of sparse sequences, implying that it is easier to recover the pulse shape in most cases.\par 

Finally, we also compared the overall success rates of the samplers at different SNR levels in Fig. \ref{overall_success_rates_vs_SNR}. Considering the recovery of sparse sequences, given in the left figure, all samplers perform quite similarly for lower SNRs, and the success rates increase as SNR increases, except for classical BGS. Its success rate reaches a stable level at around 20 dB and then starts slightly decreasing after 24 dB. This seems non-intuitive but at high SNRs, peaks of the likelihood function get sharper, and it becomes overwhelmingly difficult to escape from a locally optimum configuration. The proposed samplers and 3-Tuple BGS are not affected by this effect due to their improved ability to escape local optimums. The figure illustrates the success rate curves for both $\tau = 0.01$ and $\tau = 0.1$. It can be observed that at high SNRs, the number of simulations of NIGS-2 with resulting NMSE less than $\tau = 0.01$ is more than those of 3-Tuple BGS both with NMSE less than $\tau = 0.1$ and $\tau = 0.01$. This suggests that NIGS-2 is more successful at escaping local optimums. However, 3-Tuple BGS outperforms all other samplers at lower SNRs. The right figure in Fig. \ref{overall_success_rates_vs_SNR} demonstrates the same analyses for the recovery of the pulse sequence. As expected, at lower SNRs, the baseline samplers achieve higher success rates due to their extensive feasible regions. On the other hand, they are being outperformed by the proposed samplers at high SNRs, because of enhanced convergence characteristics of NIGS-1 and NIGS-2. We also note that it is possible to observe a similar performance reduction effect at high SNRs.

\section{Conclusion}\label{conclusion}
In this paper, we studied the problem of sparse blind deconvolution under a Bayesian framework and presented efficient MCMC based estimation methods for jointly recovering two unknown sequences from their noisy convolutions. We derived two different hierarchical Gibbs samplers under a NIG prior enforcing sparsity from, which forms a continuous valued alternative to the conventional BG model. While the first sampler follows a classical Gibbs sampler structure, the second one employs a PCG sampler scheme, which incorporates additional sampling steps to reduce the statistical dependence of the variables in an attempt to enhance the convergence rate. By moving the problem into a completely continuous valued framework, we avoided the computational burdens due to the discrete nature of the BG model. The proposed samplers were evaluated on an empirical basis via extensive numerical simulations and compared with the existing sampling schemes that are based on the BG model. The obtained results demonstrated the effectiveness of the proposed samplers on achieving successful recovery under various different settings. Comparisons with the baseline samplers demonstrated a significant increase in the convergence rate, along with considerable computational gains. As a result, the proposed methods can be used in a variety of real-life applications involving blind deconvolution problems with sparsity and time/frequency domain constraints.

\bibliographystyle{IEEEtran} 
\bibliography{References}

\end{document}